\documentclass[]{aa}  

\usepackage{xcolor}

\usepackage{graphicx}
\usepackage{amsmath,amssymb}
\usepackage{txfonts}
\begin{document}

   \title{Clumping and X-Rays in cooler B supergiant stars}

   \author{M. Bernini-Peron\inst{\ref{inst:ari},\ref{inst:ov}}
          \and 
          W.L.F.\ Marcolino\inst{\ref{inst:ov}}
          \and
          A.A.C.\ Sander\inst{\ref{inst:ari}}
          \and
          J.C.\ Bouret\inst{\ref{inst:cnrs}}
          \and
          V.\ Ramachandran\inst{\ref{inst:ari}}
          \and 
          J.\ Saling\inst{\ref{inst:hits}}
          \and
          F.R.N.\ Schneider\inst{\ref{inst:hits},\ref{inst:ari}}
          \and
          L.M.\ Oskinova\inst{\ref{inst:up}}
          \and 
          F.\ Najarro\inst{\ref{CAB}}
          }

   \institute{Zentrum f{\"u}r Astronomie der Universit{\"a}t Heidelberg, Astronomisches Rechen-Institut, M{\"o}nchhofstr. 12-14, 69120 Heidelberg \label{inst:ari}\\
        \email{matheus.bernini@uni-heidelberg.de}
   \and
        Observat{\'o}rio do Valongo, Universidade Federal do Rio de Janeiro, Ladeira do Pedro Ant{\^o}nio 43, CEP 20080-090, Rio de Janeiro \label{inst:ov}
    \and
        Aix-Marseille Univ, CNRS, CNES, LAM, Marseille, Francd \label{inst:cnrs}
    \and
        Heidelberger Institut f{\"u}r Theoretische Studien, Schloss-Wolfsbrunnenweg 35, 69118 Heidelberg, Germany \label{inst:hits}
    \and
        Institut f\"{u}r Physik und Astronomie, Universit\"{a}t Potsdam, Karl-Liebknecht-Str. 24/25, D-14476 Potsdam, Germany\label{inst:up}
    \and
   {Departamento de Astrof\'{\i}sica, Centro de Astrobiolog\'{\i}a, (CSIC-INTA), Ctra. Torrej\'on a Ajalvir, km 4,  28850 Torrej\'on de Ardoz, Madrid, Spain\label{CAB}}
    }

   \date{Received March 21, 2023; accepted June 23, 2023}

    \abstract   
    {B supergiants (BSGs) are evolved stars with effective temperatures between $\sim$$10$ to $\sim$$30$\,kK. Knowing the properties of these objects is important to understand massive star evolution. Located on the cool end of the line-driven wind regime, the study of their atmospheres can help us to understand the physics of their winds and phenomena such as the bi-stability jump.
    }
    {Despite being well-studied stars, key UV features of their spectra have so far not been reproduced by atmosphere models for spectral types later than B1. In this study, we aim to remedy this situation
    by performing quantitative spectral analyzes that account for the effects of X-rays and clumping in the wind. In addition, we also briefly investigate the evolutionary status of our sample stars based on the stellar parameters we obtained.
    } 
    {We determined photospheric and wind parameters via quantitative spectroscopy using atmosphere models computed with CMFGEN and PoWR.
    These models were compared to high-resolution UV and optical spectra of four BSGs: HD206165, HD198478, HD53138, and HD164353. We further employed GENEC and MESA tracks to study the evolutionary status of our sample.
    }
    {When including both clumping and X-rays, we obtained a good agreement between synthetic and observed spectra for our sample stars. For the first time, we reproduced key wind lines in the UV, where previous studies were unsuccessful. To model the UV spectra, we require a moderately clumped wind ($f_{\mathrm{V}\infty} \gtrsim 0.5$). We also infer a relative X-ray luminosity of about $10^{-7.5}$ to $10^{-8}$, which is lower than the typical ratio of $10^{-7}$. Moreover, we find a possible mismatch between evolutionary mass predictions and the derived spectroscopic masses, which deserves deeper investigation as this might relate to the mass-discrepancy problem present in other types of OB stars.
    }
    {Our results provide direct spectroscopic evidence that both X-rays and clumping need to be taken into account to describe the winds of cool BSGs. However, their winds seem to be much less structured than in earlier OB-type stars. Our findings are in line with observational X-rays and clumping constraints as well as recent hydrodynamical simulations. The evolutionary status of BSGs seems to be diverse with some objects potentially being post-red supergiants or merger products. The obtained wind parameters provide evidence for a moderate increase of the mass-loss rate around the bi-stability jump.
    }


   \maketitle

\section{Introduction}
\label{sec:intro}

Understanding massive stars ($M_{\rm ini} \gtrsim 8$ $\mathrm{M_\odot}$) is paramount to comprehend the visible Universe. Throughout the cosmic history, these objects inject energy, momentum, and processed material into the interstellar medium (ISM) of their host galaxies, both through their powerful stellar winds and their violent deaths, for example, as core-collapse supernovae \citep[e.g.,][]{Smith14}.
The evolution of massive stars not only determines their final fate but also the impact they have on their surrounding environment. Therefore, the quantitative study of the properties and behaviors of massive stars in different stages gives us important constraints and insights on their still quite uncertain evolution -- especially regarding post-main-sequence evolution \citep[e.g.,][]{Martins13}.

B-type supergiants stars (BSGs) are particularly  interesting as they mark an evolved stage of massive stars beyond the zero-age main sequence (ZAMS) where both core-H and core-He burning objects can be found \citep[e.g.,][]{Kraus15, Martin18}. Moreover, stars in the BSG regime may evolve bluewards or redwards \citep{Georgy14, Haucke18}. Determining the parameters of BSGs could be useful in distinguishing between these different stages and investigating the connection between OB main-sequence stars, BSGs, and other evolved objects such as yellow and red supergiants and hypergiants, Wolf-Rayet stars, and luminous blue variables \citep{Crowther06, Clark12}.

The winds of BSGs are thought to be radiatively driven, similar to their hotter O-star counterparts. However, particular phenomena inherent to their parameter space, such as an observed sharp decrease in the terminal velocities $\varv_\infty$ \citep{Lamers95} and superionization \citep[e.g.,][]{Walborn87}, turned out to be challenges for their theoretical understanding. Following early simulations by \citet{Pauldrach90} that revealed a bi-stability of solutions for the B1 hypergiant P\,Cyg, the term ``bi-stability jump'' has been introduced to refer to the sudden decrease in $\varv_\infty$. Subsequently, Monte Carlo simulations by \citet{Vink99} found a corresponding increase in the mass-loss rate\footnote{Another jump was found at even cooler temperatures in the A-star regime, which is why the bi-stability jump in the B-star regime is also referred to as the ``first'' bi-stability jump.}, giving this jump an important evolutionary relevance. However, newer modeling efforts yield conflicting results \citep[cf.][]{Krticka21,Bjoerklund22}, challenging the paradigm of a mass-loss increase for cooler B stars. 

Using the stellar atmosphere code CMFGEN \citep{Hillier98}, \cite{Crowther06} and \cite{Searle08} modeled in detail the UV and optical spectra of 31 galactic supergiants, covering the spectral types B0 to B5. Although they managed to obtain physical properties for their sample, they reported problems in reproducing important UV profiles (e.g., \ion{N}{V}, \ion{C}{IV}, and \ion{Al}{III}). Both studies did not include wind clumping and X-ray emission in their modeling, thereby leaving behind open questions about the actual wind structure and some derived properties. 

More recent papers, such as those of \cite{Martins15} and \cite{Puebla16}, analyzed BSGs including the aforementioned effects and were able to achieve better spectral fitting in the UV. Nevertheless, those papers studied only B0- and B0.5-type supergiants, which seem to adjust to the expected behavior of late-O supergiants (OSGs) with regard to their clumping and X-ray properties, namely: highly inhomogeneous winds and high relative X-ray luminosity ($L_{\mathrm{x}}/L$) \citep[][]{Driessen19, Berghoefer97}.

For these stars, the value of $\log (L_{\mathrm{x}}/L) \sim -7$ is typically employed in their modeling \citep[e.g.]{Bouret12,Puebla16}. These values are within the observational constraints of \cite{Berghoefer97}. However, for BSGs later than B1, \citeauthor{Berghoefer97} reports no detection of emitted X-rays, having only upper limits available with significant scattering. For O supergiants, \citet{Nebot-Oskinova+18} found a large spread of $\log (L_\mathrm{x}/L)$, from $-5$ to $-7.5$, raising doubt whether the value of $-7$ is universally applicable to supergiants. Moreover, \citet{Naze09} reported X-ray detections for some cool BSGs with $\log (L_\mathrm{x}/L) > -7$, albeit using only bolometric corrections based on spectral type. Nonetheless, these different findings reveal that hot supergiants in general still lack constraints on their X-ray emission properties.

In this work, we perform detailed atmosphere modeling of cool BSGs (later than B1) including both clumping and X-rays. Using both high-resolution UV and optical spectra, we analyze a small sample of BSGs in detail, namely: HD206165 (B2Ib), HD198478 (B2.5Ia), HD53138 (B3Ia), and HD164353 (B5Ib/II). In our study, we employed the CMFGEN atmosphere code to obtain the stellar and wind parameters. Additionally, motivated by \cite{PrinjaMassa10} and \cite{Petrov14}, we also model HD53138 using PoWR \citep{Graefener02,Hamann03,Oskinova07} to include an approximate treatment of optically thick clumps, and test whether such ``macroclumping'' can improve the agreement of lines such as \ion{Si}{IV} $\lambda$1394-1403 and H$\alpha$.

The rest of the paper is structured as follows: Sects.\,\ref{sec:observation} and \ref{sec:modelling} present our methodology, first explaining how we obtained and treated our observational data and then detailing how we modeled the atmosphere and the output spectra in order to obtain the relevant physical parameters. In Sect.\,\ref{sec:wind_analysis} and \ref{sec:evol_discuss} we present our results, the former focusing on the impact of clumping and X-rays in our sample stars, and the latter as an additional discussion of their evolutionary status. As closure, we draw our conclusions in Sect.\,\ref{sec:conclusion}.

%--------------------------------------------------------------------
\section{Observations and data reduction}
\label{sec:observation}

For our study, we used publicly available data of a sample of cool BSGs, presented in Table~\ref{tab:sample}. We used (i) high-resolution UV and optical spectra to obtain the wind and photospheric physical properties; (ii) magnitudes from UV to IR; and (iii) parallax distances (more details below) to estimate the stellar luminosities by reproducing the spectral energy distribution (SED).

\begin{table}
\caption{\label{tab:sample}Sample of BSGs analyzed in this paper.}
\centering
\begin{tabular}{ccccc}
\hline\hline
Star     & Name        & Sp. type & V   & distance   \\
--       & --          & --                           & mag &  kpc \\
\hline
HD206165 & 9 Cep       & B2 Ib                        & 4.76  & 0.990\tablefootmark{a}                    \\
HD198478 & 55 Cyg      & B2.5 Ia                      & 4.81  & 1.840\tablefootmark{a}                   \\
HD53138  & o$^2$ CMa   & B3 Ia                        & 3.02  & 0.847\tablefootmark{b}                   \\
HD164353 & 67 Oph      & B5 Ib/II                     & 3.93  & 0.377\tablefootmark{b}                   \\
\hline
\end{tabular}
\tablefoot{
Spectral types are retrieved from \cite{Crowther06} and \cite{Searle08}.\\
\tablefoottext{a}{Gaia eDR3 \citep{Bailer-Jones+2021}.}
\tablefoottext{b}{Hipparcos \citep{vanLeeuwen07}.}
}
\end{table}

The UV spectra were taken with IUE, using the long-wavelength (LWP) and short-wavelength (SWP) spectrographs, with the $R \sim$ 0.2 \AA\, resolution (echelle grating and cross-disperser). This is the region where most of the information on the wind is encoded for OB stars. We also used FUSE spectra available from the MAST database to better constrain the SED in the UV. 

\begin{table}
\caption{\label{tab:obs_spec_UV}IUE and FUSE spectra used in this work.}
\centering
\begin{tabular}{cccc}
\hline
\hline
HD206165     & HD198478     & HD53138 & HD164353 \\
\hline
\multicolumn{4}{c}{IUE SWP (s = small aperture)} \\
\hline
sp07365      & sp13907      & sp30153 & sp04267  \\
sp01815s     & sp36937      & sp30169 & sp0185s  \\
sp01826s     & sp36938      & sp30180 & sp02369s \\
sp02392s     & sp38687      & sp30186 & sp04369s \\
             & sp38688      & sp30202 & sp06332s \\
             &              & sp30217 &          \\
             &              & sp30223 &          \\
             &              & sp30240 &          \\
             &              & sp30247 &          \\
             &              & sp30255 &          \\
             &              & sp30264 &          \\
             &              & sp30270 &          \\
\hline
\multicolumn{4}{c}{IUE LWP}                      \\
\hline
lp06357      & lp16279      & lp10043 & lp07298  \\
             & lp16280      & lp10054 & lp07304  \\
             & lr07301      & lp10062 & lp08836  \\
             &              & lp10068 &          \\
\hline
\multicolumn{4}{c}{FUSE}                         \\
\hline
g9321601000  & p2350201000  &         &         \\
\hline
\end{tabular}
\end{table}

\begin{figure}
    \resizebox{\hsize}{!}{\includegraphics{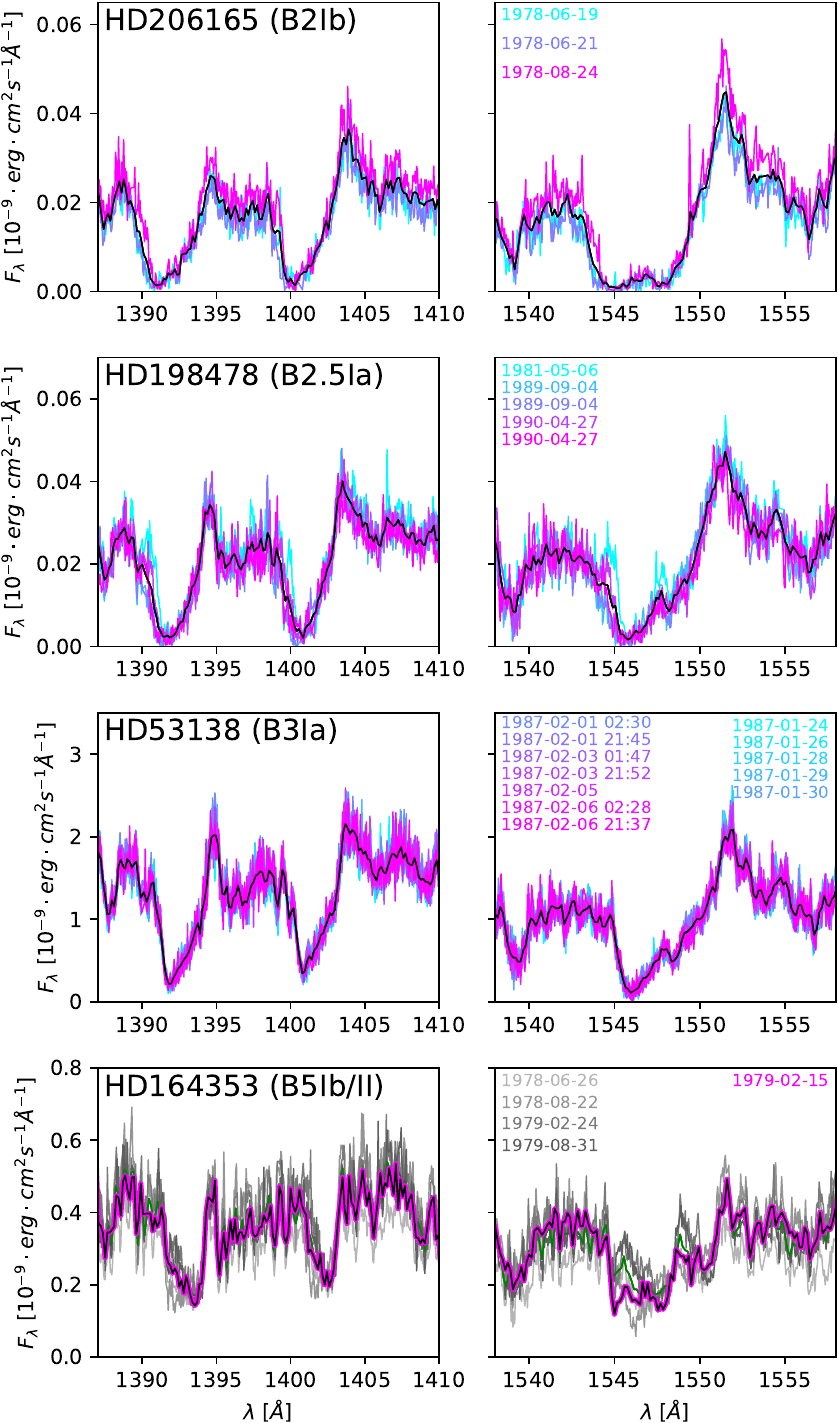}}
    \caption{UV spectra of our sample stars - CIV 1550 and Si IV 1400 region. Note: variability is absent or mild in the available IUE data as the profiles do not present drastic changes (e.g. appearance/destruction of P-Cygni profile or variations in the width). The black profiles are the computed average IUE spectra which were used for the model comparison. For HD164353, the green lines represent small aperture IUE data (re-scaled to match the flux of the large aperture acquisition). The black thick lines are the averaged spectra which were used for model comparison (see Sect.~\ref{sec:modelling}).}
  \label{fig:UV_variations}
  \end{figure}

In order to have a better signal-to-noise ratio and clearer line profiles in the UV we computed an average spectrum built from all available IUE spectra for our sample stars with high resolution and large aperture (i.e., spectra with \texttt{HIGH} and \texttt{LARGE} flags at the MAST archive). When not available, we used the spectra with the small aperture setting (spectra flagged with \texttt{SMALL}), as the absolute flux calibration is not important for the line profiles evaluation. Albeit this blurs any imprints of variability, the broad structures of the overall wind profiles (which are the focus of our study) do not change drastically over time (e.g. saturating and de-saturating or changing width noticeably), as we can see in Fig.~\ref{fig:UV_variations}. 
The star with the highest observed variability is HD164353, which may suggest that the BSG was subjected to some variation on the wind properties\footnote{Interestingly, though, the H$\alpha$ profiles of this star, which were more recently acquired, barely changed over a decade.}.
Nevertheless, such behavior does not negatively influence the focus of this study, namely the wind analysis and the obtained general stellar properties.
The spectra were manually normalized by fitting cubic splines in order to remove any spurious oscillations.
For computing the SED we used the LWP spectra as well as large-aperture spectra, regardless of the resolution. Table~\ref{tab:obs_spec_UV} lists the file number for each of the spectra obtained from the MAST archive.

The optical spectra were acquired from the POLARBASE database\footnote{\url{http://polarbase.irap.omp.eu/}} \citep{Donati97,Petit14} -- sourcing from ESPaDOnS (\texttt{R} = $\lambda/\Delta \lambda \sim 70000$) and NARVAL (\texttt{R} $\sim 65000$) -- and the ESO Archive -- using HARPS (\texttt{R} $\sim 120000$) and FEROS (\texttt{R} $\sim 48000$) spectra. For each star and instrument, we selected the spectra with the highest signal-to-noise ratio.
These spectra, which practically cover the whole optical region (3900 to 7000 \AA), were normalized with the same procedure as the UV spectra. 

\begin{figure}
    \resizebox{\hsize}{!}{\includegraphics{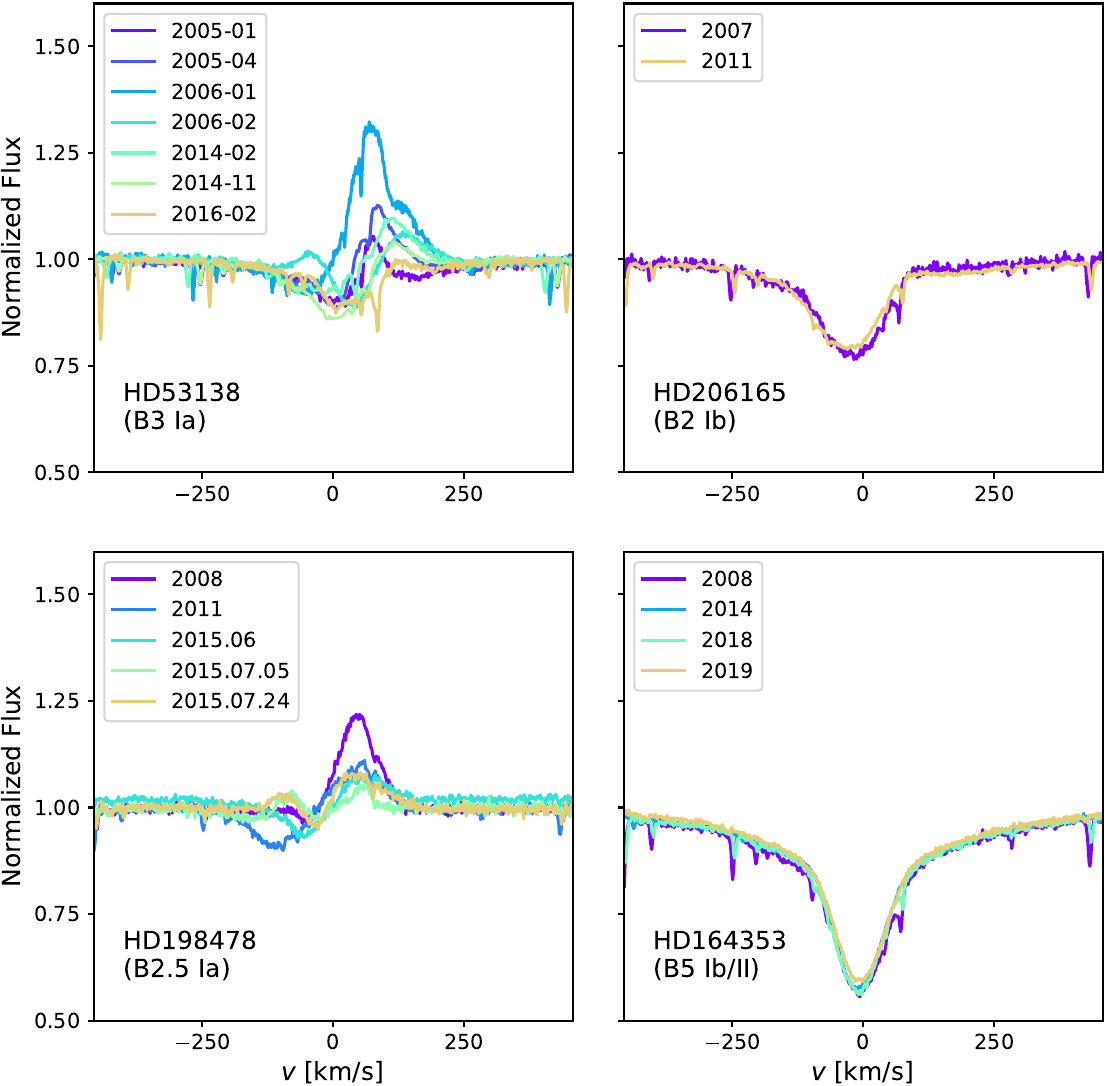}}
    \caption{Time sequences of H$\alpha$ profiles for our sample BSGs. The high variability in HD53138 and HD198478 makes the profile not very suitable for obtaining reliable wind properties without auxiliary diagnostics.}
  \label{fig:Halpha}
  \end{figure}

The BSGs in our sample show consistent spectra over time and a ``well behaved'' SED\footnote{This can be verified at the \textit{VizieR Photometry viewer} -- \url{http://vizier.u-strasbg.fr/vizier/sed/}.}. This means that the photospheres of these stars are not subject to significant large-scale variability that would spoil determinations of the main properties by modeling their spectra.
The only major variability in the spectra of these stars are seen in the H$\alpha$ profiles, which significantly vary over the years as illustrated in Fig.\,\ref{fig:Halpha}. This variability is more pronounced in  HD198478 (B2.5Ia) and HD53138 (B3Ia).

For HD198478, \citet{Kraus15} also tracked the variations of the H$\alpha$ profile for about one month and showed that its variability happens quite quickly. They concluded that this is unlikely to reflect sharp mass-loss rate variations. Instead, \citet{Kraus15} suggested large-scale inhomogeneities propagating through the wind, reinforcing the importance of including clumping in the analysis of these objects. For HD53138, the H$\alpha$ variability was studied by \citet{Morel04}, who reported no periodicity in the variability behavior. \cite{Martin18} pointed out this variability could be due to the presence of magnetic fields, but to date, the literature offers no evidence for their presence in the star's atmosphere.

The photometric magnitudes were obtained from multiple sources: FUV and NUV are taken from GALEX \citep{BeitiaAntero16}, UBVI from the XHIP catalog \citep{Anderson12} and R is obtained from \cite{Morel78}; Gaia's Gbp, G, and Grp were taken from Gaia eDR3 \citep{Fabricius21}; JHK were obtained from 2MASS \citep{Cutri03}; and WISE's W1 and W2 were obtained from the AllWISE catalog \citep{Cutri13}. 

For HD198478 and HD206165, we use the distances from \citet{Bailer-Jones+2021} employing Gaia eDR3 parallaxes.
The distances for HD53138 and HD164353 were derived from the revised Hipparcos parallax \citep{vanLeeuwen07}. We opted not to use Gaia's parallax for these stars because they are considerably bright (V = 2.94 and 3.96 respectively), which is close to the allowed limit of Gaia \citep{Lindegren18}. As \citet[][Sect.~3]{Golovin23} also discusses, for such bright stars the Hipparcos parallaxes tend to be more reliable. Moreover, the ``renormalized unit weighted error'' (\texttt{ruwe}) for these stars amounted to 3.62 and 2.51, values that are much greater than the cut-off of 1.4 suggested by \cite{Fabricius21}. 

\section{Stellar atmosphere models}
\label{sec:modelling}

In order to model the BSG atmospheres and analyze their spectra, we used the stellar atmosphere codes CMFGEN \citep[][]{Hillier98, Hillier03} and PoWR \citep{Graefener02,Hamann03,Sander15}. Both codes compute the radiative transfer in a co-moving frame formalism and solve the statistical (non-LTE) and radiative equilibrium equations in an expanding atmosphere assuming spherical symmetry and stationarity. They also account for the line-blanketing effect arising from the millions of line transitions in heavier elements such as iron. Once the atmospheric stratification is converged, the synthetic spectrum is computed via a formal solution of the radiative transfer in the observer's frame \citep[see][for details]{Mihalas78}. While the codes have very different architectures, which might produce small variations of inferred parameters, the underlying physics within each are similar and throughout the past years were used to successfully model spectra of hot stars \citep[cf.][for a comparison in modeling $\zeta$ Pup as an example]{Oskinova07,Bouret12}.

In this work, we did not solve the equation of motion consistently as we aim to obtain empirically derived parameters for the wind properties. Instead, as typically done in the literature, a fixed wind velocity structure was adopted in the form of a (modified) $\beta$-law  \citep[see, e.g.,][]{Bouret12}, which smoothly connects the wind regime with the quasi-hydrostatic inner layers. The connection velocity in our models is set to $10\,\mathrm{km}\,\mathrm{s}^{-1}$, slightly lower than the sound speed.

The winds of hot massive stars are line-driven and inherently unstable \citep{Lucy70}. Therefore, time-dependent models predict contrasts in terms of density, velocity, and temperature that could give rise to X-ray emission in the outflow \citep[see, e.g.,][]{Owocki88}. These effects can have an impact on the emerging spectra and have been extensively studied over the last decades \citep[see, e.g.,][and references therein]{Bouret05, Sundqvist14}. In the following sections, we present how the effects of clumping and X-rays are taken into account in the CMFGEN and PoWR models.

\subsection{Clumping and X-Rays in CMFGEN}

CMFGEN is able to account for wind inhomogeneities via the ``microclumping'' formalism, where the wind is considered to consist only of optically thin clumps with a void interclump medium. In addition, the clump sizes are assumed to be smaller than the photons' mean free path. The volume filling factor is defined as
\begin{equation}
\label{eq:clumping}
    f_\text{V}(r) = f_{\text{V},\infty} + (1 - f_{\text{V},\infty})\,e^{-\varv(r)/\varv_\mathrm{cl}}
,\end{equation}
where $f_{\text{V},\infty}$ is a free parameter denoting the value at $r \rightarrow \infty$. $\varv(r)$ is the velocity at radius $r$ and $\varv_\mathrm{cl}$, the second free parameter, is a characteristic velocity describing the onset of clumping (i.e., where it starts to become relevant). 
A newer version of CMFGEN presented by \citet{Flores21} allows the inclusion of clumping with arbitrary optical depths, which provides a description with fewer assumptions on the inhomogeneities. However, only a few objects had been analyzed in such a way and this version is not currently public.

X-ray emission can be included in CMFGEN as a product of thermal \textit{Bremsstrahlung} following a similar approach as \cite{Baum92} and \cite{Pauldrach94}. This formalism follows from the paradigm that X-rays are produced by shocks within the wind, aiming to parameterize this type of emission.
As \cite{Hillier98} explains, the X-ray is added during the calculation of the opacities as an additional emissivity term.
In the version used in this paper (2017 release) the X-ray emissivities are further making use of the APEC table \citep[][]{Smith01} for solar metallicity to account for X-ray line opacities\footnote{The CMFGEN manual states that the non-inclusion of the emissivity table would severely underestimate the derived X-ray luminosity.}. In the current description, three parameters describe the wind-intrinsic X-rays, namely: (i) the X-ray filling factor $f_\mathrm{x}$, which can be interpreted as the amount of shocks producing X-rays in the wind, (ii) $T_\mathrm{x}$, which represent the temperature of the shocks, defining the X-ray distribution in frequency space, and (iii) the X-ray characteristic velocity $\varv_\mathrm{x}$, which determines the spatial distribution of the  X-ray emission. From this parametrization, the code computes the wind emitted and observable X-ray luminosity $L_\mathrm{x}$ (with energies > 0.1 keV) as an output.

\subsection{Clumping and X-Rays in PoWR}
\label{sec:cl_xr_powr}

Similarly to CMFGEN, PoWR allows the inclusion of clumping with the same description. In addition, PoWR has the option to include an approximated formalism for optically thick clumping in the formal integral \citep{Oskinova07}. In this so-called ``macroclumping'' formalism, the effective opacity, $\kappa_{\mathrm{eff}}$, is written as:
\begin{equation}
    \label{eq:macroclump}
    \kappa_{\mathrm{eff}}(r) = \kappa (r) \left ( \frac {1 - e^{-\tau_C(r)}}{\tau_C(r)} \right),
\end{equation}
where $\kappa$ is the opacity considering microclumping only.
$\tau_C(r)$ describes the optical depth of the clumps and is defined as 
\begin{equation}
    \tau_C(r) = \kappa(r)L_0 \left (\frac{r^2}{R_*^2 f_\mathrm{V}^2(r)}\frac{\varv(r)}{\varv_\infty} \right )^{1/3},
\end{equation}
where $L_0$ denotes the typical separation between the clumps.

Concerning shock-generated X-rays parametrization, PoWR includes these as emissions from thermal \textit{Bremsstrahlung} as described in \cite{Baum92}. While this is similar to CMFGEN, the main difference is that PoWR so far does not include any emissivity tables -- see \ref{app:xray_params} for more details on the difference in implementation between codes. Consequently, the effect of X-rays is treated as additional continuum emissivities only. However, this difference is not important for the present work, as we employ the PoWR models only to study the impact of macroclumping on specific line profiles (see Sect.\,\ref{ssc:macroclumping}).

\subsection{Atomic Physics}

For the model atoms in our CMFGEN models, we included 16 different elements with a total of 46 different ions. This marks a considerable extension compared with \cite{Crowther06} and \cite{Searle08}, who used 11 elements and 27 ions. Moreover, we extended the number of so-called ``superlevels'', a concept introduced by \cite{Anderson89, Anderson91} to combine several similar-energy levels into a single ``superlevel'' and reduce the number of equations that need to be solved for the statistical equilibrium. We summarize our underlying atomic data in Table \ref{tab:atomic}. 
A more detailed description of the atomic data and their implementation in CMFGEN can be obtained from the CMFGEN manual\footnote{\url{http://kookaburra.phyast.pitt.edu/hillier/web/CMFGEN.htm}} as well as \cite{Hillier98} and \cite{DessartHillier10}.

\begin{table}
\caption{\label{tab:atomic}Ions and number of levels, superlevels and transitions considered per ion used in our B supergiant models..}
\centering
\begin{tabular}{lccc}
\hline
\hline
Ion    & levels & superlevels & transitions \\
\hline
H I    & 30                         & 30                              & 435                                                    \\
He I   & 69                         & 69                              & 905                                                    \\
He II  & 30                         & 30                              & 435                                                    \\
C II   & 39                         & 21                              & 202                                                    \\
C III  & 243                        & 99                              & 5528                                                   \\
C IV   & 64                         & 64                              & 1446                                                   \\
N I    & 104                        & 44                              & 855                                                    \\
N II   & 105                        & 44                              & 898                                                 \\
N III  & 287                        & 57                              & 6223                                                   \\
N IV   & 70                         & 44                              & 440                                                    \\
N V    & 49                         & 41                              & 519                                                \\
O I    & 199                        & 58                              & 4193                                                   \\
O II   & 137                        & 340                             & 8937                                                   \\
O III  & 104                        & 36                              & 761                                                    \\
O IV   & 56                         & 32                              & 359                                                    \\
Ne II  & 48                         & 14                              & 328                                                    \\
Ne III & 71                         & 23                              & 460                                                \\
Mg II  & 45                         & 18                              & 362                                                    \\

Mg III & 201                        & 29                              & 3052                                                   \\
Al II  & 58                         & 36                              & 270                                                    \\
Al III & 80                         & 80                              & 2011                                                   \\
Si II  & 80                         & 52                              & 628       \\
Si III & 147                        & 99                              & 1639      \\
Si IV  & 66                         & 66                              & 1090      \\
P IV   & 90                         & 30                              & 656       \\
P V    & 62                         & 16                              & 561       \\
S III  & 44                         & 24                              & 193       \\
S IV   & 142                        & 51                              & 1503      \\
S V     & 101                        & 40                              & 831        \\  
Ca III & 90                         & 30                              & 868       \\
Cr II  & 1000                       & 84                              & 66400     \\
Cr III & 1000                       & 68                              & 73962     \\
Cr IV  & 234                        & 29                              & 6354      \\
Mn II  & 1000                       & 58                              & 49066     \\
Mn III & 1000                       & 47                              & 70218     \\
Mn IV  & 464                        & 39                              & 19176     \\
Fe II  & 827                        & 62                              & 13182     \\
Fe III & 607                        & 65                              & 6670      \\
Fe IV  & 1000                       & 100                             & 37899     \\
Fe V   & 1000                       & 139                             & 37737     \\
Ni II  & 1000                       & 59                              & 33555     \\
Ni III & 150                        & 24                              & 1345      \\
Ni IV  & 200                        & 36                             & 2337   \\
Ni V   & 183                        & 46                              & 1524    \\ 
\hline
\end{tabular}
\end{table}

\subsection{Stellar parameter determination}

In order to obtain the physical properties, we aim to reproduce all the diagnostic lines in the UV and optical regimes simultaneously. This procedure consists of the following major steps:

\paragraph{Effective temperature and surface gravity:} We determine the effective temperature $T_\mathrm{eff}$ using the ionization balance of \ion{Si}{ii} $\lambda${4128-30}, \ion{Si}{III} $\lambda\lambda${4552-68-75}, as well as \ion{He}{I} $\lambda${4471} and \ion{Mg}{II} $\lambda${4481}. Given that \ion{Si}{IV} $\lambda\lambda${4089,4116} appears weakly for most of the targets we also use them as auxiliary $T_\mathrm{eff}$ diagnostics. For the surface gravity $\log g$, we use the wings of H$\beta$, H$\gamma$, and H$\delta$. The typical errors are 1\,kK and 0.1\,dex following a conservative estimate -- similarly to \citet{Crowther06} and \citet{Searle08} -- and to account for uncertainties due to microturbulence, which affects most of the optical lines, especially \ion{Si}{III} $\lambda\lambda${4552-68-75}.
    
\paragraph{CNO abundances:} Once these main properties are established, we derive the C, N, and O abundances using mainly \ion{C}{II} $\lambda${4267}, \ion{N}{II} $\lambda${4447}, \ion{N}{II} $\lambda\lambda${4601--4630} and \ion{O}{II}, $\lambda${4069}, \ion{O}{ii} $\lambda${4590--96,4661}. The typical error is 0.3 dex, as we did not perform any fine-tuning -- in contrast, for example, to what \citet{Puebla16} did for HD37128 (B0Ia). The other elemental abundances are assumed to be solar \citep[][]{Asplund09} and the He/H number fraction is assumed to be 0.1. This value represents a slightly enriched abundance (corresponding to a mass fraction of $X_\text{He} = 0.28$) since it is expected that BSGs are evolved massive stars with some surface He enrichment \citep{Maeder09}. Determining the abundance of He is not trivial and, according to \citet{Crowther06}, would not produce significant differences in the lines.

\paragraph{Line broadening:} The rotational and macroturbulent broadening for our sample, $\varv \sin i$ and $\varv_\mathrm{mac}$ respectively, are taken from the IACOB survey \citep{SimonDiaz17}\footnote{The determined values for several stars from this study can be found at: \url{https://vizier.cds.unistra.fr/viz-bin/VizieR?-source=J/A+A/597/A22}.}. The values are shown in Table~\ref{tab:phot_prop}.

\paragraph{Luminosity and extinction:} The luminosity, $L$, is then obtained by reproducing the SED from the UV to the infrared, directly using the flux-calibrated IUE spectra and the flux derived from the available photometry and applying the corresponding distances and the reddening, which is obtained by fitting the 2175\,\AA\ absorption bump. The extinction was determined following the reddening law from \citet{Cardelli89} with the values obtained for each star shown in Fig~\ref{fig:SED}. For HD53138 and HD164353, whose distances came from Hipparcos \citep{vanLeeuwen07}, we consider an error in $L$ of 35\%, which is the largest error in the distance for this part of the sample. Since these two objects have low reddening, the uncertainties associated with $E(B-V)$ are negligible. Conversely, for HD198478 and HD206165, whose distances are sourced in Gaia eDR3, we have an error of $\lesssim 12\%$. However, as these objects are subject to higher extinction ($E(B-V) \sim 0.5$), we consider a $L$ uncertainty of 20\% in order to account for that. 
  
\paragraph{Masses and radii:} Using $L$, $\log g$, and $T_{\mathrm{eff}}$ we can obtain the masses ($M$) and stellar radii ($R$). We considered that the uncertainties in the three aforementioned parameters follow a Gaussian distribution centered on the determined values, with their error estimates as the standard deviations (see Appendix \ref{app:statistics} for details). From that, we computed the mass probability distribution for each star. To estimate the uncertainty in mass, we take the standard deviations of these resulting distributions. We obtained an error of 40\% in the masses for the stars whose distances are based on Gaia parallaxes and 50\% for those using Hipparcos data. Concerning the radii, we obtained an error of 15\% and 20\% respectively.

\subsection{Wind parameter determination}

\begin{figure}
    \resizebox{\hsize}{!}{\includegraphics{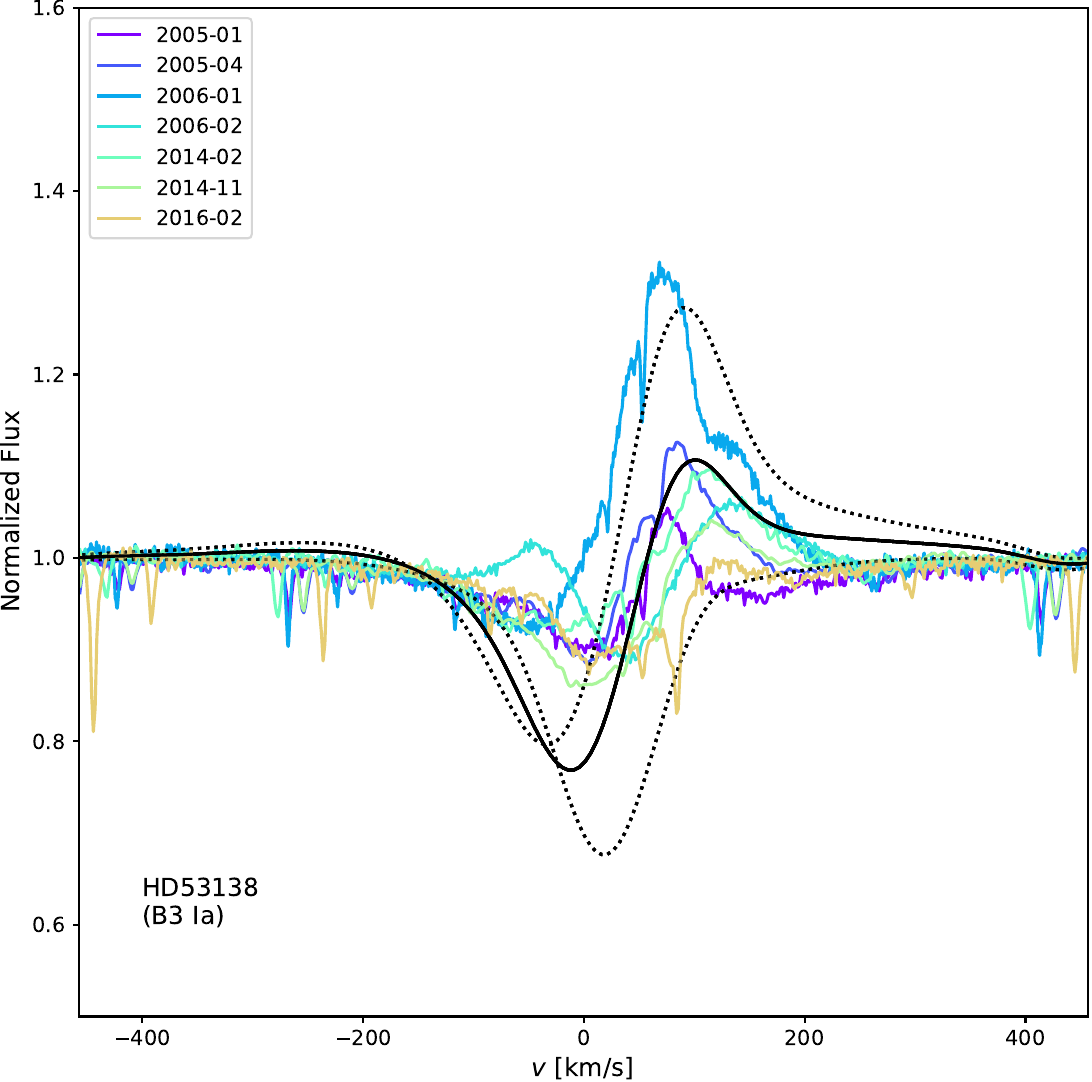}}
    \caption{Comparison between synthetic and observational spectrum of HD53138 in the H$\alpha$ region. The color scheme is the same as in Fig.~\ref{fig:Halpha} while the black filled line represents the preferred model while the and dotted lines represent models with an increase/decrease in $\dot{M}$ by a factor of 30\%.} 
  \label{fig:Halpha2}
  \end{figure}

After determining the photospheric parameters, we then obtained the wind properties, using mainly the UV diagnostics:

\paragraph{Mass-loss rates:} We infer the clumping-corrected mass-loss rate ($\dot{M}/\sqrt{f_{\text{V},\infty}}$) from the H$\alpha$ strength and the ``main'' UV P-Cygni profiles, namely: \ion{N}{V} $\lambda${1238-42}, \ion{C}{II} $\lambda${1335-36}, \ion{Si}{IV} $\lambda${1394-1403}, \ion{C}{IV} $\lambda${1548-50} and \ion{Al}{III} $\lambda${1855-63}.
Due to the variability in H$\alpha$, we did not attempt to fit this line precisely. Instead, we focus on obtaining models with compatible H$\alpha$ strengths, namely, allowing flux discrepancies not greater than 10\% relative to the observations (see Fig.\,\ref{fig:Halpha2} for an illustrative example).

Thus, an error of 30\% in $\dot{M}$ was adopted based on how the synthetic spectra start to clearly deviate from the observations. 
    
\paragraph{Wind velocity:} The velocity gradient parameter $\beta$ is fixed at 2.0 based on the value obtained by \citet{Crowther06} and on the fact that according to \citeauthor{Crowther06} and \citet{Searle08} a higher $\beta$ is preferred in the case of cooler BSGs\footnote{Although \citet{Searle08} adopt $\beta = 1.5$ for HD53138.}. Moreover, a more recent study on BSGs conducted by \citet{Haucke18}, who analyzed the optical spectra of 19 stars using FASTWIND \citep{Puls05} and found $\beta = 2.0$ for HD53138 and a systematic trend of $\beta > 2.0$ for other cool BSGs. The terminal velocities were initially taken from \citet{Howarth97}. However, in order to properly fit the main UV lines' absorption widths, we adjusted them for each star if necessary. The adopted error associated with $\varv_\infty$ is 100 km\,s$^{-1}$. 

\paragraph{Microturbulence scaling:} The microturbulence in the photosphere $\xi_{min}$ was constrained using the depth of \ion{Si}{III} $\lambda\lambda${4552-68-75} and photospheric UV absorption lines. In CMFGEN, the microturbulence $\xi_{turb}$ in the formal integral is assumed to grow monotonically outward until it reaches its maximum value $\xi_{max}$, which is assumed to be $0.1\varv_\infty$ as commonly done in the literature, following 
    \begin{equation}
        \xi_\mathrm{turb} = \xi_\mathrm{min} + \frac{\xi_\mathrm{max} - \xi_\mathrm{min}}{\varv_\infty}.
    \end{equation}
    
\paragraph{Wind inhomogenieties:} Regarding clumping, the characteristic velocity $\varv_\mathrm{cl}$ is fixed to $30$\,km\,s$^{-1}$ while assuming that clumping starts already at the innermost layers of the wind -- for all models, as it is commonly done in OB star analyses \citep[e.g.,][]{Marcolino09, Puebla16}. Moreover, \cite{Torrejon15} analyzed X-ray emission in QV Nor, a B0I + neutron star (NS) binary and provided empirical evidence that clumping starts at $r < 1.25 \mathrm{R_*}$, namely, very close to the photosphere. This would correspond to low wind velocities, in line with our assumption. For estimating the volume filing factor ($f_{\mathrm{V},\infty}$), we tested models with different values of $f_{\mathrm{V},\infty} = 0.1$, $0.2$, $0.5$, and $1.0$, the lattermost corresponding to a smooth wind,  always keeping $\dot{M}/\sqrt{f_{\mathrm{V},\infty}}$ constant.

\paragraph{X-rays:} For the X-ray parameters we keep the characteristic velocity fixed at around 0.7$\varv_\infty$. This choice is motivated by the fact that shock-induced X-rays would require a high dispersion of velocities in the wind and thus could only emerge further away from the star where the wind achieved higher speeds ($\sim$$400$ km\,s$^{-1}$). In O- and early BSGs, such values can be reached closer to the photosphere, $\sim 0.2 \varv_\infty$ \citep{Lagae21}. In Appendix\,\ref{app:xray_params} we discuss how this manifests in the spatial distribution of the X-ray emission of our models. This is also in line with expectations from hydrodynamical simulations by \citet{Driessen19}.

Once $\varv_\mathrm{x}$ is fixed, we test how different combinations of $T_\mathrm{x}$ and $L_\mathrm{x}$ lead to a good enough fitting of the UV wind lines. Using HD53138 (B3 Ia) as the primary target, we performed a detailed investigation of the possible X-ray parameters before testing the validity of the results on the other stars of the sample.

\section{Wind clumping and X-ray results}
\label{sec:wind_analysis}

\begin{figure}
  \resizebox{\hsize}{!}{\includegraphics{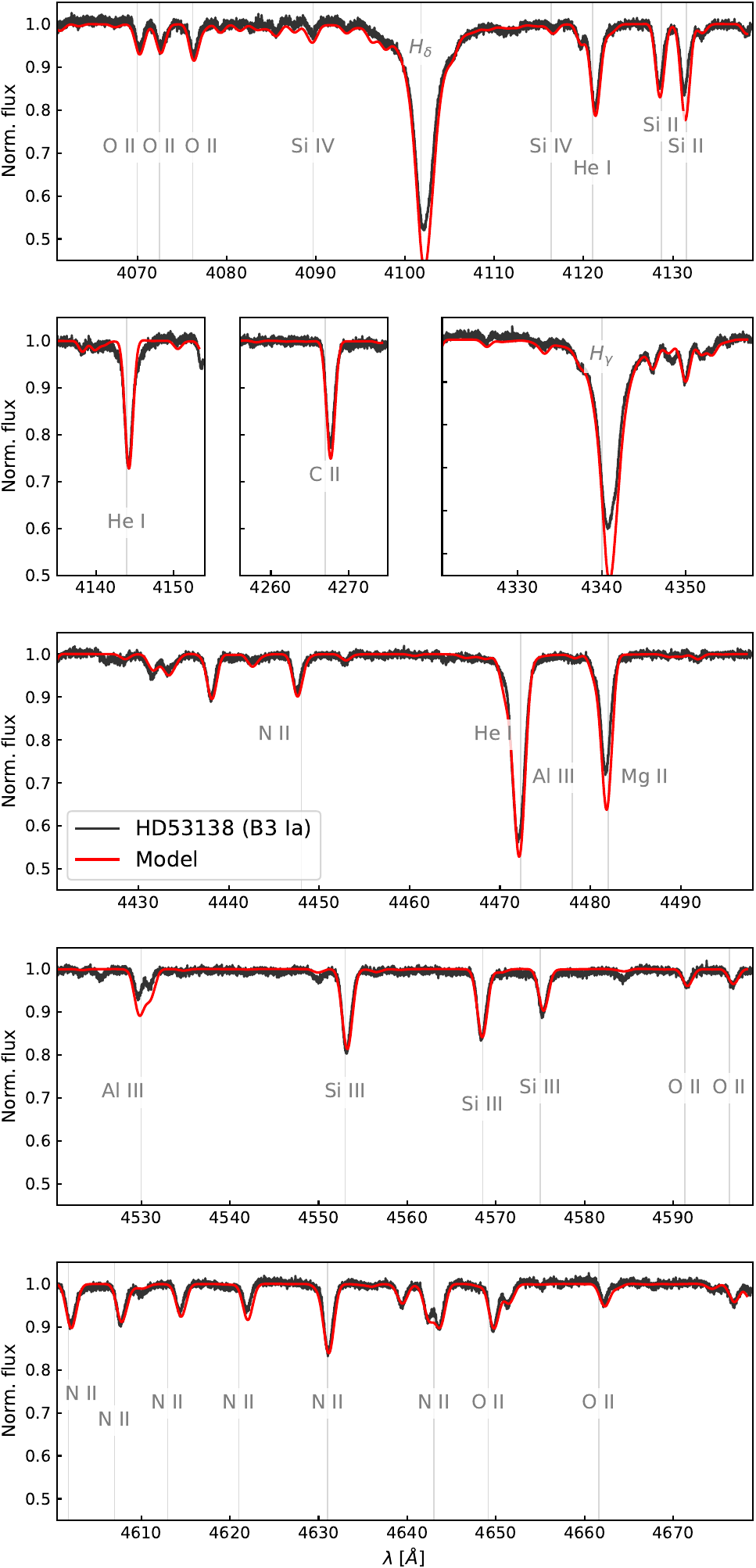}}
  \caption{Optical spectrum of HD53138 (black solid line) compared to our best-fit model (red-solid line). Here we depict the main diagnostic lines used to obtain the stellar photospheric properties.}
  \label{fig:diagnostics}
\end{figure}

By fitting the observed UV and optical spectra of the sample stars, we obtained the main stellar properties. Using HD53138 as a showcase, we depict the best fit of the main (photospheric) diagnostic lines in Fig.\,\ref{fig:diagnostics}. From the figure, we can notice that a satisfactory fitting is achieved for most of the lines, especially the diagnostic features.

The main stellar properties of our sample BSGs are presented in Table~\ref{tab:phot_prop} and the results of the spectral fitting can be found in Appendix~\ref{app:master_plots}. Additional information about the individual BSGs can be found in Appendix \ref{app:info_other_BSGs}. In Table~\ref{tab:prop_HD53138}, we compare our results for HD53138 with various studies from the literature, generally finding a good agreement despite the partially different underlying methodologies. A more in-depth discussion about the evolutionary status and consequences of the stellar properties obtained is undertaken later in Sect.\,\ref{sec:evol_discuss}. % The 

% STELLAR PHOTOSPHERIC PROPERTIES

\begin{table*}
\caption{\label{tab:phot_prop}Main stellar parameters of our sample BSGs. The uncertainties are discussed in Sect.\,\ref{fig:diagnostics}. The solar values are in the table only for comparative purposes.}
\centering
\begin{tabular}{llcccc|c}
\hline
\hline
Star & & HD206165 & HD198478 & HD53138 & HD164353 & Sun (A09) \\
Sp. Type&  & B2Ib & B2.5Ia & B3Ia & B5Ib/II & G2V \\
\hline
$T_{\mathrm{eff}}$ & $\mathrm{[kK]}$ & 18.0$\pm1.0$ & 16.0$\pm1.0$ & 16.0$\pm1.0$ & 15.0$\pm1.0$ & 5.8 \\
$\log g$ & ${\rm [cgs]}$ & 2.45$\pm0.15$ & 2.15$\pm0.15$ & 2.15$\pm0.15$ & 2.50$\pm0.15$ & 4.43 \\
$\log (L/\mathrm{L_\odot})$ & -- & 5.07$\pm0.10$ & 5.57$\pm0.10$ & 5.14$\pm0.15$ & 4.05$\pm0.15$ & 0.0 \\
$M$ & $\mathrm{[M_\mathrm{\odot}]}$ & 13.0$\pm5.2$ & 33.0$\pm12.8$ & 12.0$\pm6.0$ & 2.8$\pm1.4$ & 1.0 \\
$\xi_{\mathrm{min}}$ & ${\rm [km\,s^{-1}]}$ & 15 & 17 & 20 & 23 & \\
$\varv \sin i$ & ${\rm [km\,s^{-1}]}$ & 39 & 38 & 38 & 25 &  \\
$\varv_\mathrm{mac}$ & ${\rm [km\,s^{-1}]}$ & 59 & 54 & 56 & 52 &  \\
$\mathrm{X_C}$ & -- & -3.47$_{(7.60)}$ & -3.29$_{(7.78)}$ & -3.29$_{(7.78)}$ & -3.14$_{(7.93)}$ & -2.62$_{(8.43)}$ \\
$\mathrm{X_N}$ & -- & -3.00$_{(8.00)}$ & -2.61$_{(8.40)}$ & -2.68$_{(8.32)}$ & -2.61$_{(8.40)}$ & -3.15$_{(7.83)}$ \\
$\mathrm{X_O}$ &  -- & -2.43$_{(8.52)}$ & -2.71$_{(8.23)}$ & -2.58$_{(8.36)}$ & -2.41$_{(8.53)}$ & -2.24$_{(8.69)}$ \\
\hline
\end{tabular}
\tablefoot{CNO values are given in the logarithm of the total mass fraction. The subscripts show the corresponding chemical abundance values in units of $12 + \log (n_{\mathrm{C,\,N,\,O}}/n_{\mathrm{H}})$. A09 stands for \cite{Asplund09}.}
\end{table*}

%%%%%%%%%%% HD53138 PROPERTIES COMPARISON
\begin{table*}
\caption{\label{tab:prop_HD53138}Comparison of the main stellar parameters of HD53138 (B3Ia) with different studies in the literature.}
\centering
\begin{tabular}{lcccccccc}
\hline \hline
$T_{\mathrm{eff}}$  & $\log g$ &  $\log L$     & $R_\star$                    & $M$  & C   & N                    &     O           & Source    \\ 
$\mathrm{[kK]}$    & ${\rm [cgs]}$ & -                     & $\mathrm{[R_\mathrm{\odot}]}$   & $\mathrm{[M_\mathrm{\odot}]}$   & -- & -- & -- \\
\hline
18000 & 2.2  & 5.32                 & 46                   & 12 &             --             &           --          &            --              & H18            \\
17000 & 2.1  & 5.2                  & 71                   & 16 &           --               &        --              &               --           & M18, Z09 \\
15400 & 2.15 &       --               &         --              & 24 &         --                 & 8.22                 &             --             & F10            \\
16500 & 2.25 & 5.3                  &       55               & 20 &                   {7.78} & 8.29                 &           {8.75}          & S08       \\
15500 & 2.05 & 5.34                 & 65                   & 17 &                    {7.95} & 8.45                 &             {8.15}       &  C06         \\
\hline
16000 &   2.15   &  5.14               & 49               &   12   & 7.78 & 8.32 & 8.36 &                                                                   This work \\
\hline
\end{tabular}
\tablefoot{C06 stands for \cite{Crowther06}, S08 for \cite{Searle08}, F10 for \cite{Fraser10}, Z09 for \cite{Zorec09}, M18 for \cite{Martin18}, and H18 for \cite{Haucke18}.}
\end{table*}

%%%%%%%%%%%%%

After the determination of the photospheric properties, we analyze the winds via the UV spectra of our sample stars. A particular emphasis here is to reproduce the so-called ``superionization'', a phenomenon denoting the presence of highly ionized elements which are not expected from the temperature regime reflected by $T_\text{eff}$ \citep{Cassinelli-Olson79}. 

In previous works by \cite{Crowther06} and \cite{Searle08}, the biggest problem in reproducing the UV spectrum occurred due to the presence of both \ion{C}{II} $\lambda$1335-36 and \ion{C}{IV} $\lambda$1548-50 with the former typically showing an overly strong P\,Cygni profile while the latter did hardly show any wind signature. 

\begin{figure}
    \resizebox{\hsize}{!}{\includegraphics{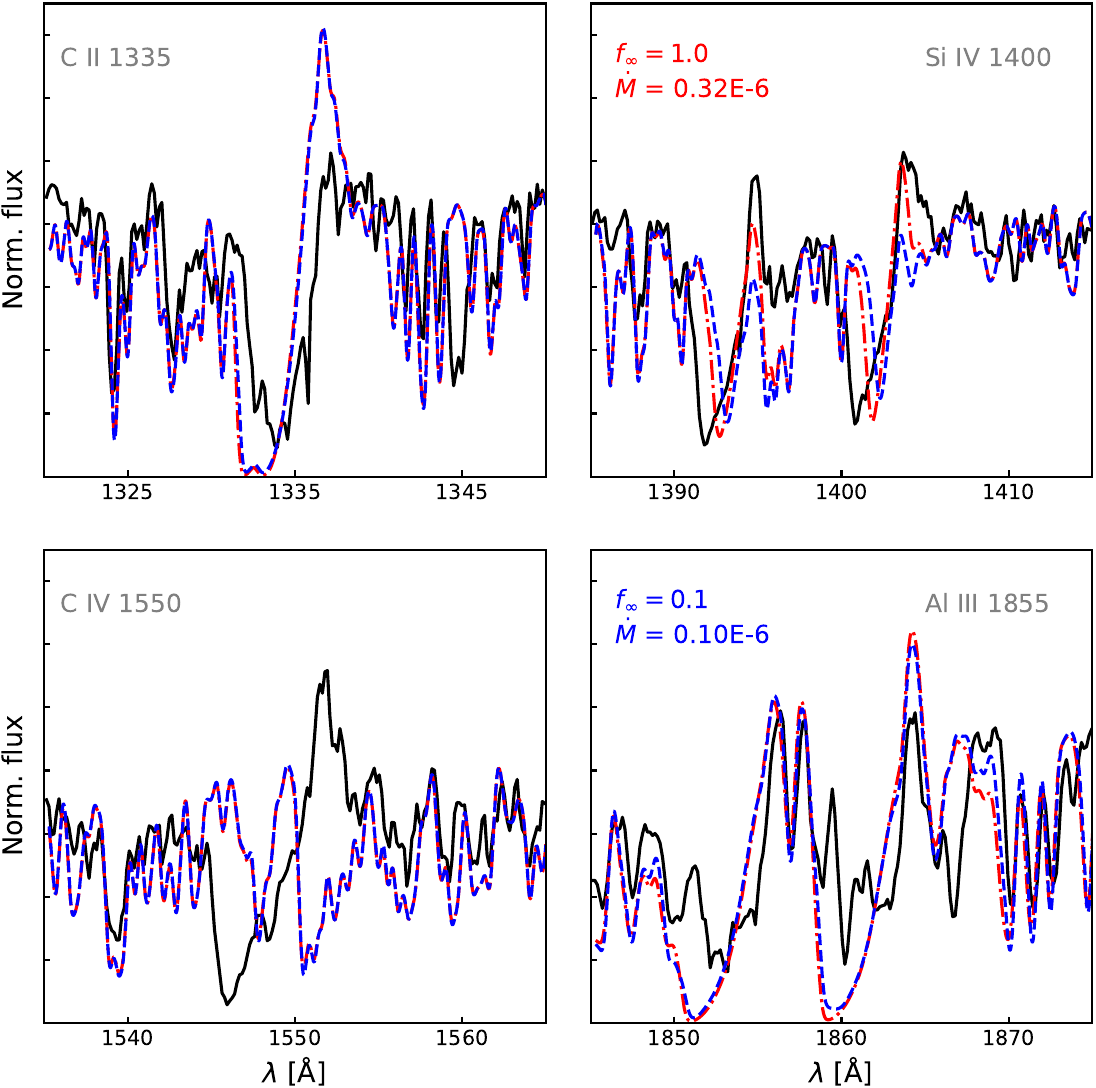}}
    \caption{Comparison between models (red dash-dotted and blue dashed lines) and observed (black full line) spectra of HD53138. The red model includes neither clumping nor X-rays in the wind and illustrates the current status in the literature regarding the UV spectral fitting of cool BSGs. The blue model includes clumping ($f_{\mathrm{V},\infty} = 0.1$) but no X-rays.}
  \label{fig:noXR_model}
  \end{figure}

We illustrate this in Fig~\ref{fig:noXR_model}, where we depict a model for HD53138 without clumping and X-rays, reproducing the findings of \citeauthor{Searle08} and \citeauthor{Crowther06}. In the same figure, we also show that clumping alone is not enough to address the observed superionization.

\begin{figure}
    \resizebox{\hsize}{!}{\includegraphics{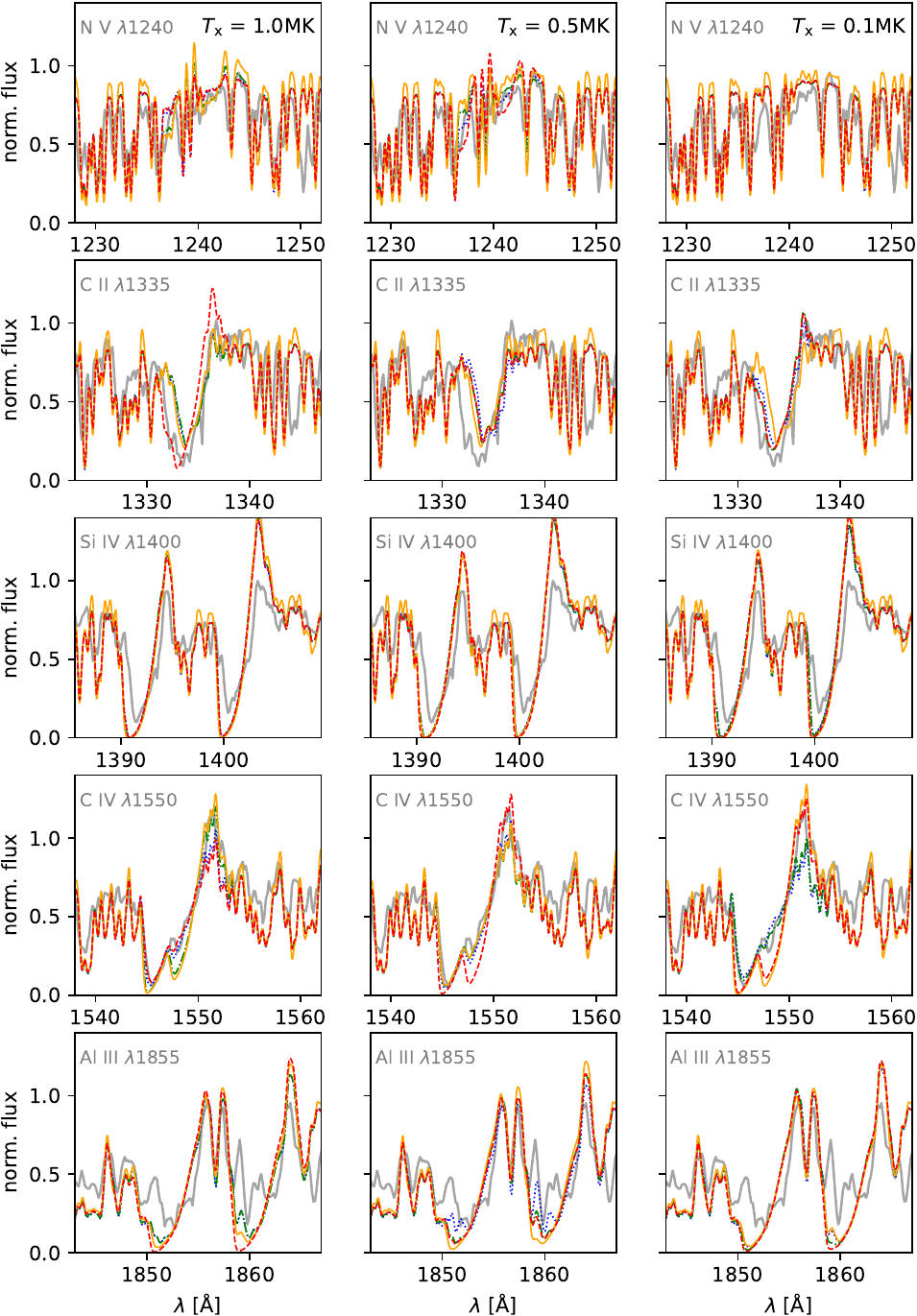}}
    \caption{Best fitting models for HD53138 in the UV region, focusing on the main P-Cygni profiles, namely \ion{N}{V} $\lambda$1238-42, \ion{C}{II} $\lambda$1335-36, \ion{Si}{IV} $\lambda$1394-1403, \ion{C}{IV} $\lambda$1548-50 and \ion{Al}{III} $\lambda$1855-63, depicted in each row. Each column of panels show model with the same shock temperature. For each panel the red-dashed stands for $f_\infty = 1.0$, orange/full for $f_\infty = 0.5$, green-dash-dotted for $f_\infty = 0.2$ and blue-dotted for $f_\infty = 0.1$.}
  \label{fig:dog_UV_fitting}
  \end{figure}

In Fig.\,\ref{fig:general_UV_fitting}, we show our new results for HD53138, where we tested different combinations of clumping and X-ray parameters. As illustrated, we achieve a good agreement for the two carbon as well as most of the other UV lines, even when varying some of the X-ray and clumping parameters. The overall good agreement in Fig.\,\ref{fig:dog_UV_fitting} has already indicated that, despite the general lack of clear X-ray detections, the inclusion of X-rays in modeling cool BSG atmospheres appears to be vital. A more detailed view is shown in Appendix~\ref{app:UV_fits}. 

However, with the present X-ray implementation, there is no clear preference for a particular X-ray parametrization, as the fit quality for all lines is somewhat similar. We therefore present and take into account different solutions obtained in our analysis instead of presenting only one, singular solution. Among those, we assign formally one of them as preferred to guide some of our later discussions.
Concerning \ion{Si}{IV} $\lambda$1394-1403, the model profiles are always stronger than observed and even saturated in the models. Given that \citet{Crowther06}
and \citet{Searle08} obtained weaker lines than in the observations, we can conclude that this line is highly affected already by a moderate amount of X-rays: For every $T_\mathrm{x}$ and $f_{\text{V},\infty}$ tested, \ion{Si}{IV} $\lambda$1394-1403 becomes a strong P-Cygni profile. To avoid saturation, the resulting $L_\mathrm{x}$ would need to be reduced so much, that it would spoil the fitting of \ion{C}{IV} $\lambda$1548-50.

For \ion{Al}{III} $\lambda$1855-63, we obtain a similar overprediction, but with an opposite trend. The profile only de-saturates at higher $L_\mathrm{x}$. This is in line with the overprediction of \ion{Al}{III} $\lambda$1855-63 reported by \citeauthor{Crowther06} and \citeauthor{Searle08} when using models without X-rays.

Due to the conflicting behavior of \ion{Si}{IV} and \ion{Al}{III}, we put more emphasis on the reproduction of the C profiles. A more in-depth discussion of the Si lines in the context of optically thick clumping is given in Sect.\,\ref{ssc:macroclumping}.
Exhibiting the broadest disagreement in previous studies, the carbon profiles provide a better idea of the overall ionization structure of the cool BSG winds as both profiles belong to the same element. Therefore, we consider a model as ``good'' if it recovers both lines simultaneously as unsaturated (albeit not too weak) P\,Cygni profiles.

Another line to consider is \ion{N}{V} $\lambda$1238-42: Because of its very weak profile, we also considered models which do not predict \ion{N}{V} at all, but still reproduce both C lines adequately, as potential solutions. Whether or not \ion{N}{V} $\lambda$1238-42 is taken into account has direct consequences on the predicted X-ray properties for these stars.

\subsection{Limits of the X-ray luminosity}
  \label{sec:xraylimits}

For HD53138, we are able to find a simultaneous fit of \ion{C}{II} $\lambda$1335-36 and \ion{C}{IV} $\lambda$1548-50 for models with $\log (L_\mathrm{x}/L) \sim -13$ and $T_{\mathrm{x}} = 0.1$\,MK (largely independent\footnote{A model without clumping yields a good fit for \ion{C}{II} $\lambda$1335-36 and \ion{C}{IV} $\lambda$1548-50 with $\log (L_\mathrm{x}/L) \sim -13.3$.} of the adopted clumping). Such a value of $T_\mathrm{x}$, representing a shock temperature ($T_\mathrm{shock}$), is roughly aligned with the relation 
\begin{equation}
T_\mathrm{shock} \sim \left(\frac{\Delta \varv}{300\,\mathrm{km\,s^{-1}}}\right)^2 \cdot 10^6\,\mathrm{K},    
\end{equation}
between $T_\mathrm{shock}$ and the wind velocity dispersion ($\Delta \varv$) obtained by analysing O and early-B stars in \cite{Cohen14}
if we considers a wind velocity dispersion of $\sim 70$ km\,s$^{-1}$ \citep[cf.\ the BSG model results by][]{Driessen19}.
However, none of the models with $T_{\mathrm{x}} = 0.1$\,MK (and the resulting very low $L_\mathrm{x}$) were able to yield any trace of \ion{N}{V} $\lambda$1238-42, which seems to appear to be a very weak profile in the observed spectrum of HD53138 and in the other sample BSGs as well.
In short, this means that if we cannot detect \ion{N}{V} $\lambda$1238-42 in the spectrum of cool BSGs, $\log (L_\text{x}/L)$ as low as $\sim -13$ cannot be discarded.

For $T_{\mathrm{x}} = 0.5$\,MK, both C lines as well as \ion{N}{V} $\lambda$1238-42 can be reproduced simultaneously. In the solution with no clumping ($f_{\text{V},\infty}=1.0$), we require a lower $\log (L_\mathrm{x}/L) \approx -8.3$, whilst for very clumped winds we obtain values up to $\log (L_\mathrm{x}/L) \approx -7.5$ for $f_{\text{V},\infty}=0.1$. 
 
Reasonable spectral fits are also obtained when adopting $T_{\mathrm{x}} = 1.0$\,MK. However, this temperature is hard to motivate as a shock temperature \citep[e.g.,][]{Cohen14, Driessen19}. For HD53138 with $\varv_\infty = 680$ km\,s$^{-1}$, a velocity dispersion of $\sim 300$ km\,s$^{-1}$ would be required, which seems implausible. Later-type BSGs in general show even lower wind velocities.
Thus, taking into account the wind-shock paradigm, our model set with $T_{\mathrm{x}} = 0.5$\,MK would be considered as the preferred description. The good fits with $T_{\mathrm{x}} = 1$\,MK would instead require a different theoretical justification. Among the $T_{\mathrm{x}} = 0.5$\,MK models, the solutions with  $f_{\text{V},\infty}=0.2$ and $0.5$ are scarcely different. For our later comparison discussions, we pick the $0.5$ model as our preferred solution. The model yields a slightly stronger \ion{C}{ii} profile and thus is marginally better in comparison to the observations.

Combining all model efforts we can conclude that a minimum $\log (L_\text{x}/L) \approx -8.3$ is required to reproduce the UV spectrum of HD53138. This is in line with the current observational upper limit of $-7.4$ \citep{Berghoefer97}.

We applied the same X-ray parameters -- allowing for small modifications such as $\varv_\mathrm{x}$ and $\log (L_\mathrm{x}/L)$ for better reproducing the spectra when necessary -- for the other sample BSGs, considering also the same set of $f_{\mathrm{V},\infty}$. The resulting spectral fits for the other stars are compiled in Fig.\,\ref{fig:general_UV_fitting_other_BSGs}. 

We can see that the lines are reproduced reasonably well for all our BSGs, exhibiting a behavior that is similar to that of HD53138 with respect to the reproduction of the UV P-Cygni profiles. While this as such already represents a significant improvement compared to the previous studies, there are also specific results worth commenting on for each of the targets:

\paragraph{HD206165:} For this object, \ion{C}{IV} $\lambda$1548-50 was successfully recovered as a P-Cygni profile and best reproduced regardless the $T_{\mathrm{x}}$ by the models with lower clumping. This is expected given the higher $T_\mathrm{eff}$ and the scaling requirement that smoother winds imply a higher $\dot{M}$.
While we also reproduced the P-Cygni profile of \ion{C}{II}, contrary to \citep{Searle08}, we could not find a solution that  matches the full width of its absorption trough. This implies that the population of \ion{C}{II} is underestimated in the outer wind region of our models. Given the higher effective temperature of HD206165 compared to HD53138, this is more difficult to achieve in the atmosphere modeling framework applied in this work.
However, the fitting of \ion{Si}{IV} and \ion{Al}{III} yield  good results, as the observed lines are close to saturation. For all other sample stars, the predicted \ion{Al}{III} $\lambda$1855-63 is stronger than observed.
Among the solutions, only those with $T_\mathrm{x} = 1.0$~MK and $f_\mathrm{V,\infty} = 1.0$ or 0.5 yield a a meaningful \ion{N}{V}. While a high $T_\mathrm{x}$ is usually not preferred within the wind-shock paradigm, this BSG has the highest $\varv_\infty$ ($900$~$\mathrm{km\,s^{-1}}$) in our sample. While the solutions for $f_\mathrm{V,\infty} = 1.0$ and $0.5$ are generally indistinguishable, we formally selected the model with $f_\mathrm{V,\infty} = 1.0$, as our preferred solution due to its slightly stronger \ion{N}{V} profile.
    
\paragraph{HD198478:} Similarly to HD206165, this star was best fitted by models with lower clumping, as more inhomogeneous wind models predict overly weak C lines. Furthermore, both stars have very strong \ion{Si}{iv}, also largely recovered by our models. Still, one can notice a slight oversaturation, similarly to HD53138. The profile starts to de-saturate when $f_{\text{V},\infty}$ is decreased, but for the models with $T_\mathrm{x} = 0.5$~MK, they immediately fail to produce a P-Cygni for \ion{C}{II} and predict an overly strong \ion{N}{V}. For the case with $T_\mathrm{x} = 1.0$~MK, \ion{C}{IV} is not fitted at all with lower $f_{\text{V},\infty}$. Thus, among the solutions, the model with $f_{\text{V},\infty} = 1.0$ and $T_\mathrm{x} = 0.5$~MK is the preferred option.
  
\paragraph{HD164353:} For our target with the latest spectral type, all models reproduce \ion{C}{II} quite well, but only the models with lower clumping yield acceptable \ion{C}{IV} profiles. For $T_\mathrm{x} = 0.5$ MK, the model with no clumping (and, thus, the highest $\dot{M}$) produces overly strong \ion{N}{V} lines. For an increasing amount of clumping, \ion{N}{V} decreases. Thus, our model with $f_\mathrm{V,\infty} = 0.5$ yields the best fit for this line, making it our preferred solution. Models with other values of $T_\mathrm{x}$ are not sufficient as they fail to produce noticeable \ion{N}{V} profiles.

In HD164353, the observed \ion{Si}{IV} line is weaker than in our other targets, showing almost a ``photospheric'' appearance. In this case, the models with lower clumping yield  an overprediction of the \ion{Si}{IV} lines. The observed unsaturated \ion{Si}{IV} could only be reproduced with highly clumped models, implying an extremely low mass-loss rate ($\sim 10^{-9}$), which would unfortunately spoil the reproduction of the \ion{C}{IV} profile. This pattern of not being able to obtain an unsaturated \ion{Si}{IV} together with a reasonable profile for \ion{C}{IV} is similar to HD198478.
    
\paragraph{Overall X-ray limits:} From the spectral fitting of our sample stars, we can obtain estimates of the X-ray luminosities intrinsically emitted by the winds of these stars. We present them in Table~\ref{tab:LXs_allstars}, where we also show the upper limits given by \cite{Berghoefer97}.
If we allow for a very low shock temperature, which is not sufficient to produce \ion{N}{V}, we derive values as low as $\log (L_\text{x}/L) \sim -13$. Acknowledging that the \ion{N}{V} line is barely noticeable, we consider this value to be  a ``hard'' lower limit for the X-ray luminosities of these stars. However, if we take \ion{N}{V} into account, we need to consider higher $T_\text{x}$ and then we find values for $\log (L_\text{x}/L)$ between $-7.0$ and $-8.3$ for reasonable spectral fits. We thus see the latter value of $\sim -8.3$ as the more likely actual lower limit for the intrinsic $L_\text{x}/L$ ratio of cooler BSGs.

Some of our models also yield values of $\sim -7.0$, thus matching the ``canonical'' value known for hotter massive stars \citep[e.g.,][]{Sana06}. However, the results of our study do not favor these models as they require a high degree of clumping, which in general tends to weaken (and spoil) the carbon lines. Additionally, considering the whole sample, the best fits were obtained by models with $f_\mathrm{V,\infty} = 0.5$ and $1.0$ -- namely, with mild and no clumping in the wind.  Therefore, we consider lower values of clumping to be preferred for describing the winds of these cooler BSGs, which is in line with recent simulations and observational studies \citep{Driessen19, RubioDiez22}. Thus we provide additional empirical-based evidence that the winds of cool BSGs are smoother than those of O- and early BSGs.

In summary, we suggest that cooler BSGs must have X-ray luminosities that are below the ``canonical'' value of $-7.0$, namely, a value in the region between $-7.3$ and $-8.3$. These findings are in line with the observational constraints of \citet{Berghoefer97} for our sample stars.

\begin{table}
\caption{\label{tab:LXs_allstars}X-ray luminosities range for the whole sample. Boldface values highlight the preferred solutions.}
\centering
\begin{tabular}{cc|ccc}
\hline
\hline
& \rule[0mm]{0mm}{3.5mm} & $\frac{T_\mathrm{x}}{\mathrm{MK}} = 0.1$ & $\frac{T_\mathrm{x}}{\mathrm{MK}} = 0.5$ & $\frac{T_\mathrm{x}}{\mathrm{MK}} = 1.0$  \\
$f_{\mathrm{\infty}}$ & $\dot{M}$ & $\log \frac{L_\mathrm{x}}{L}$ &  $\log \frac{L_\mathrm{x}}{L}$ & $\log \frac{L_\mathrm{x}}{L}$  \\
-- & $\mathrm{M_\odot}/\mathrm{Myr}$ & --  & --  & -- \\
\hline
    \multicolumn{5}{c}{HD206165 -- $L_\mathrm{x}/L$ upper limit: -7.3 \citep{Berghoefer97}}\\
\hline
0.1 & 0.066 & -12.1 & -7.3 & -7.6  \\
0.2 & 0.090 & -13.0 & -7.4 & -8.1  \\
0.5 & 0.150 & -13.0 & -7.3 & -7.3  \\
\textbf{1.0} & \textbf{0.211} & -13.7 & -7.4 & \textbf{-8.7}  \\
\hline
    \multicolumn{5}{c}{HD198478 -- $L_\mathrm{x}/L$ upper limit: -7.1 \citep{Berghoefer97}} \\
\hline
0.1 & 0.150 & -13.0 & -8.0 & -8.0  \\
0.2 & 0.220 & -13.3 & -8.0 & -8.0  \\
\textbf{0.5} & \textbf{0.502} & -13.3 & \textbf{-8.0 }& -8.0  \\
1.0 & 0.711 & -13.4 & -8.1 & -8.1  \\
\hline
    \multicolumn{5}{c}{HD53138 --$L_\mathrm{x}/L$ upper limit: -7.6 \citep{Berghoefer97}}\\
\hline
0.1 & 0.110 & -12.7 & -7.6 & -7.0  \\
0.2 & 0.146 & -13.0 & -8.0 & -7.0  \\
\textbf{0.5} & \textbf{0.234} & -12.8 & \textbf{-8.4} & -7.5  \\
1.0 & 0.325 & -13.3 & -8.3 & -8.0  \\
\hline
    \multicolumn{5}{c}{HD164353 -- $L_\mathrm{x}/L$ upper limit: -6.9 \citep{Berghoefer97}} \\
\hline
0.1 & 0.003 & -12.7 & -7.5 & -7.0  \\
0.2 & 0.004 & -13.0 & -8.3 & -7.3  \\
\textbf{0.5} & \textbf{0.007} & -13.3 & \textbf{-8.0} & -7.3  \\
1.0 & 0.010 & -13.3 & -8.3 & -8.0  \\
\hline

\end{tabular}

\end{table}

In appearent contrast to our results, \cite{Naze09} measured X-ray (0.5 keV $<h\nu <$ 10 keV) fluxes of OB stars with XMM-Newton and detected X-rays in seven cool BSGs. For some of their targets, they obtained $\log (L_\mathrm{x}/L) \equiv \log F_x/F > -7$, where $F_x$ and $F$ are their labels for the X-ray and bolometric unreddened flux respectively. 
However, none of their targets overlap with our sample. Moreover, their sample contains colliding wind binary candidates, as well as stars that have not been thoroughly investigated yet, and their bolometric fluxes were computed solely with a standard recipe based on bolometric corrections and intrinsic colors derived from the spectral type. Thus, further studies are required to check whether all cooler BSGs really have an intrinsic $\log (L_x/L) < -7.0$. In any case, both our and their results underline that X-rays seem to be intrinsic to at least a significant fraction of the cool BSGs and need to be considered when modeling their atmospheres.

\subsection{Effects of optically thick clumping}
\label{ssc:macroclumping}

Previously, we demonstrated that the inclusion of (micro-)clum\-ping, and especially X-rays substantially improved the fitting of the UV carbon profiles for our sample stars. However, all of our solutions for the different BSGs predict too strong \ion{Si}{IV} $\lambda$1394-1403 and \ion{Al}{III} $\lambda$1855-63. Moreover, the absorption part of the derived H$\alpha$ profiles is in every case stronger than the observed spectrum, which may indicate a systematic problem in the ionization structure.

Analyzing BSG winds with the SEI method \citep{Lamers87}, \cite{PrinjaMassa10} found that the lines in the \ion{Si}{IV} $\lambda$1394-1403 doublet have optical depth ratios smaller than $2$. They concluded that the profile must be affected by optically thick clumps in the wind. In BSGs, also H$\alpha$ could behave like a resonance line, as discussed by \citet{Petrov14}. Hence, this profile could be affected by optically thick clumps as well.

Given this motivation, we tested the inclusion of macroclumping as described in \citet{Oskinova07}. In this simplified initial approach, the macroclumping plays no role in impacting the density or ionization structures. Instead, as explained in Sect.~\ref{sec:cl_xr_powr}, it changes the opacity in the computing of the observer's frame spectrum. 

Keeping the properties we obtained with our CMFGEN model for HD53138, we calculated PoWR models with and without macroclumping for this star. For this task, we selected input options for PoWR that were as close as possible to the CMFGEN description, such as the radial clumping stratification or the considered X-ray luminosity. Despite the technical differences in the codes (cf.\ Appendix\,\ref{app:xray_params}), the PoWR models produce a spectrum similar to the CMFGEN models, reflecting the similar physical framework of the two codes. However, we did not aim for a precise benchmark between the two codes. Thus, we did not perform further iterations for a fine-tuned reproduction of the observation with the PoWR models employed for this task; rather, we focused on the effect on the main wind lines instead. 

  \begin{figure}
    \resizebox{\hsize}{!}{\includegraphics{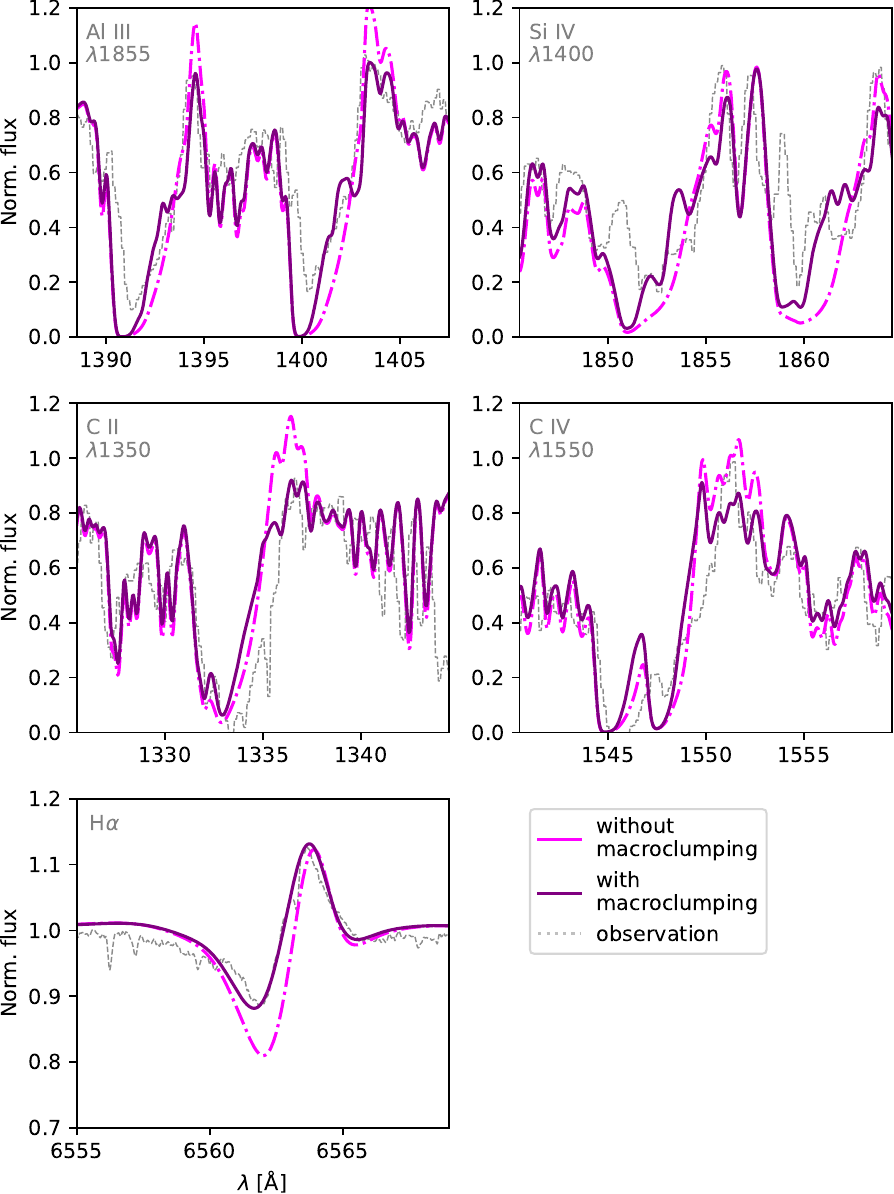}}
    \caption{Impact of the inclusion of optically-thick clumps on the resolution of the formal integral on the stronger wind profiles. Both models have the same parameters, except for the presence of macroclumping in the formal integral. For the model that includes macroclumping (purple-solid line), we use $L_0 = 0.5$ R$_\star$.}
  \label{fig:macroclump}
  \end{figure}

The synthetic spectra of our PoWR models with and without macroclumping are shown in Fig.\,\ref{fig:macroclump}. In order to obtain this outcome, we employed a typical separation between clumps $L_0 = 0.5$ R$_\star$. 
Notably, the most relevant impact of the macroclumping treatment is the decrease of the emission part of the P-Cygni profiles, except for H$\alpha$ and \ion{Al}{III} $\lambda1855$ to a small extent, since they are more affected in their absorption component.

The weakening of the absorption component in H$\alpha$ and \ion{Al}{III} $\lambda1855$ can be attributed to the reduction of the effective opacity. Due to the optically thick clumps, the number of available absorbing atoms\footnote{The number density is proportional to $1/L_0$.} is lower as a fraction of the material is shielded, and thus not contributing to the opacity of resonance lines \citep{Oskinova07}.
Likewise, the reduction in the effective opacity tends to reduce the emission components.

For \ion{Si}{IV} $\lambda$1394-1403, despite improving the fit in the emission part, the major discrepancy (i.e., the non-observed saturation of the absorption) was not substantially affected and thus it remains unresolved by this approximate macroclumping treatment. The carbon lines, especially \ion{C}{II} $\lambda$1335-36, only show minor effects in their emission parts. For these lines, as they were already closer to full saturation in the models without macroclumping, the impact of the reduction of the absorbers is limited.

In short, we obtain improved profiles for H$\alpha$ and, to a smaller degree, in \ion{Al}{III} and \ion{Si}{IV}. 
However, in order to properly quantify the impact of macroclumping, we would need to integrate the effects of optically clumps into the density and opacity structure -- for instance, an alteration in the opacity of strong profiles will affect the radiative transport within the wind, which could change the ionization structure. However, despite the approximate treatment of macroclumping, we provide evidence that optically thick clumping likely plays a role in shaping some of the wind-affected line profiles in later-type BSGs, which is in line with the findings of \citet{PrinjaMassa10} and \cite{Petrov14}.

\subsection{Mass-loss rates of cool BSGs}

%%% TABLE WIND PROPERTIES Vinf beta ksi_max Mdot f Mdot/f^(1/2) Lx/L v_x 
%TABLE WIND
\begin{table*}
\caption{\label{tab:wind_prop}Mechanical wind properties of the sample stars. For comparison, we also show literature results.}
\centering
\begin{tabular}{cccc|cccccc}
\hline
\hline
Star      & Sp. Type  & $\varv_{\infty}$     & $\log \dot{M}$  & C06 & L07 & S08 & K15 & H18 & R22 \\
   --      &    --     & km\,s$^{-1}$        & $\mathrm{M_\odot}$/yr  \\
\hline
HD206165  & B2Ib      & 900  &    \textbf{-6.67}$_{1.0}$ & -- & -- & -6.30|\textbf{{-6.37}} & -- & -- & -- \\
HD198478  & B2.5Ia    & 570    &   \textbf{-6.14}$_{1.0}$ & -6.63|\textbf{-6.19} & -- & -6.30|\textbf{-5.94} & {-6.62}|\textbf{-6.01} & -- & $\lesssim$-6.42|\textbf{$\lesssim$-5.90} \\
HD53138   & B3Ia      &  680   &   \textbf{-6.63}$_{0.5}$ & {-6.44}|\textbf{-6.63} & {-6.51}|\textbf{-6.46} & {-6.34}|\textbf{-6.46} & -- & {-6.62}|\textbf{-6.65} & $\lesssim$-5.74|\textbf{$\lesssim$-5.93} \\
HD164353  & B5Ib/II   & 640    &   \textbf{-8.15}$_{0.5}$ & -- & -- & -7.22|\textbf{-7.36} & -- & -- & -- \\

\hline

\end{tabular}
\tablefoot{Boldface values indicate the mass-loss rates re-scaled to the radii we obtained for each BSG. Subscripts indicate the chosen clumping values. C06 stands for \cite{Crowther06}, L07 for \cite{Lefever07}, S08 for \cite{Searle08}, K15 for \cite{Kraus15}, H18 for \cite{Haucke18}, and R22 for \cite{RubioDiez22}.}
\end{table*}

After deriving and discussing the clumping and X-ray properties of our sample, we now have sufficient information to compare our results with theoretical predictions and discuss potential consequences. Our wind properties are presented in Table~\ref{tab:wind_prop}. After testing different values for $f_{\text{V},\infty}$, we were able to confirm that the product $\dot{M}/\sqrt{f_{\text{V},\infty}}$ has to stay constant in order to yield good spectral fits. Thus, solutions with different $f_{\text{V},\infty}$ also have different $\dot{M}$. In Table~\ref{tab:wind_prop}, we list only the preferred solutions according to the tailored analysis described on Sect.~\ref{sec:xraylimits} with respect to the choice of volume filling factor. The $f_{\text{V},\infty}$ is indicated as a subscript to the value of $\dot{M}$.

When considering the mass-loss rates re-scaled by the respective obtained radii, our analysis in general points towards lower mass-loss rates compared to previous work. For HD206165 and HD164353, the differences are most prominent, while the results for the two other targets are within the derived uncertainties of $\dot{M}$ (about 0.2 dex).
Compared with previous works, the most significant distinction in our models are the presence of X-rays, which allowed us to successfully obtain the observed super-ionization and reproduce multiple UV wind lines -- which were used as $\dot{M}$ diagnostics. This contrasts with the previous studies in the literature which needed to rely basically only H$\alpha$ as a spectroscopic diagnostic. Additionally,  we find that clumping was not a decisive factor for influencing the derived $\dot{M}$ because we found values that would imply (almost) smooth winds, close to the description employed in previous studies \citep[e.g.,][]{Crowther06, Searle08}. Moreover, \cite{Krticka21}, who produced cool BSGs models that include clumping but not X-rays, did not reproduce higher ions such as \ion{C}{IV} (see their Fig.\,9). In that line, we also noticed that in our set of models, those without X-rays, regardless of the clumping values were not able to produce superionization.

\begin{figure}
  \resizebox{\hsize}{!}{\includegraphics{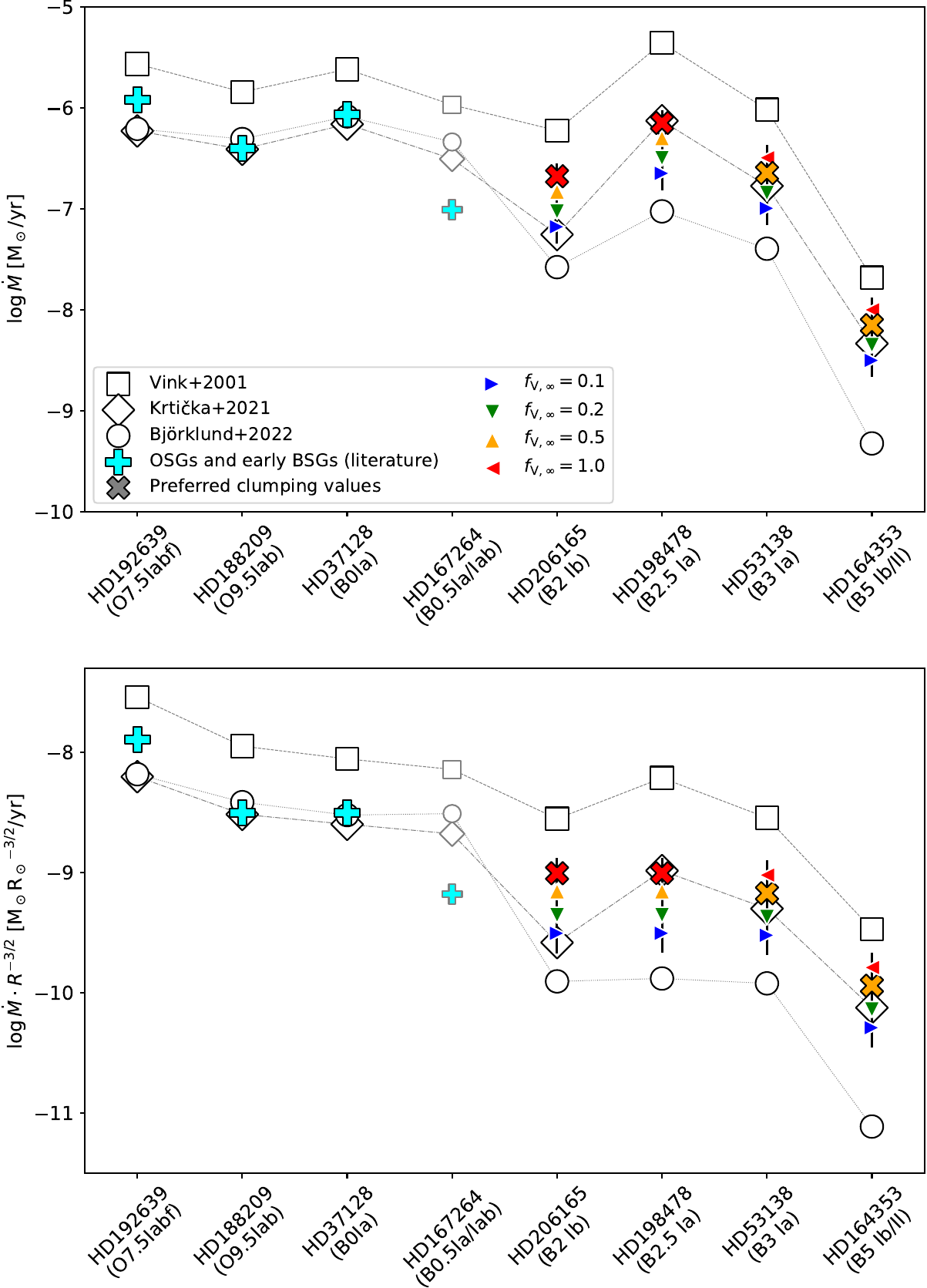}}
  \caption{Comparison between our derived mass-loss rates and the predictions according to different recipes, \citet{Vink00}, \citet{Krticka21}, and \citet{Bjoerklund22}, for our sample stars shown in the upper panel. Early BSGs and OSGs, represented with cyan crosses, which were also analyzed using clumping and/or X-rays are included, namely: HD192639 \citep[O7.5Iabf,][]{Bouret12}, HD188209 \citep[O9.5Iab,][]{Marcolino17}, HD37128 \citep[$\epsilon$-Ori, B0Ia][]{Puebla16} and HD167264 \citep[B0.5Ia/Iab,][]{Martins15}. Bottom panel shows the same as the upper panel, but with all the stars scaled to the same radii following \cite{Puls06} scaling relation of OB stars. HD167264 $\dot{M}$ values have fainter symbols in order to indicate that the parameters might not be very reliable as the star is a binary \citep{Mahy22}.}
  \label{fig:Mdot_comparison_combined}
\end{figure}

In Fig.\,\ref{fig:Mdot_comparison_combined}, we compare the derived mass-loss rates values for each BSG with the mass-loss recipes of \citet{Vink00}, \citet{Krticka21}, and \citet{Bjoerklund22}. Despite the preference for less clumped winds, we also include values of $\dot{M}$ for the other $f_\mathrm{V, \infty}$ tested as solutions with less clumped winds cannot be discarded (cf.\ Sect.~\ref{sec:xraylimits}). For comparison, we also included stars on the hot side of the bi-stability jump that were analyzed by different authors considering clumping and/or X-rays. To account for the different radii arising from different luminosities, we further compare the mass-loss rates scaled by their respective stellar radii following \citet{Puls06}.
For the cool stars, all our obtained values are located between the predictions of \citet{Vink00} and \citet{Bjoerklund22}, independent of the clumping correction. In general, the \citet{Vink00} predictions are about a factor of 2 or 3 too high, while the values from \citet{Bjoerklund22} are systematically lower than our derived results, usually by a factor of $\sim 6$. Instead, the predictions from \citet{Krticka21} fit remarkably well with our results, except for the B2Ib star HD206165. For the O- and early B- supergiants from the literature, the \citet{Vink00} discrepancy remains while the predictions of \citeauthor{Bjoerklund22} and \citeauthor{Krticka21} start to coincide and reasonably reproduce the observations, except for the B0.5 supergiant HD167264. It is important to mention though that this star is part of a binary system \citep{Sana14, Mahy22}, and the parameters derived by \cite{Martins15} were obtained without taking that into account.

The discrepancy between \citet{Krticka21} and \citet{Bjoerklund22} for cool BSGs is striking. In contrast to the results from  \citet{Krticka21} and as well as the predictions by \cite{Vink99}, the models from \citet{Bjoerklund22} do not yield any increase of $\dot{M}$ when transitioning to lower $T_\text{eff}$ values at a fixed luminosity. Empirical determinations of $\dot{M}$ \citep[e.g.,][]{Crowther06, Benaglia07, Markova08} across this bi-stability jump region (BSJR) so far could not provide clear constraints for or against a jump in $\dot{M}$, given also the spread in luminosities that complicate direct comparisons.

Our new results now prefer the \citet{Krticka21} prescription, which includes an increase in $\dot{M}$, albeit in a quantitatively different way than the \citet{Vink99} prescription commonly applied in stellar evolution models. However, the \citet{Krticka21} recipe predicts the increase to start at lower $T_\text{eff}$ than \citet{Vink99,Vink00,Vink01}, which seems to be contradicted by the B2 supergiant in our sample, which show a higher mass-loss than predicted by \citet{Krticka21}. It is evident that the biggest discrepancies with \citet{Krticka21} are approaching around the ``dividing'' line of the bi-stability jump -- $T_\mathrm{eff} \sim 22$ kK -- where supergiants of the spectral type B0.5 to B2 are located.
We thus conclude that (at least in the case of supergiants) some form of mass-loss increase seems to happen when stars evolve towards cooler $T_\text{eff}$. The precise $T_\text{eff}$ regime of this increase is hard to constrain with our limited sample but seems to be closer to the BSJR from \citet{Vink99} as well as the more recent calculations by \citet{Vink18}, albeit with an increase in $\dot{M}$ on the quantitative level found in the models from \citet{Krticka21}. A broader and more coherent sample of both hot and cool BSGs ought to be analyzed in order to draw a more conclusive picture.

Our insights on $\dot{M}$ also have potential consequences for the evolution of massive stars due to (i) the direct change of the resulting stellar mass and (ii) its impact on the loss of angular momentum \citep[see, e.g.,][]{Keszthelyi17}. A significant increase of $\dot{M}$ in the BSJR, as discussed by \citet{Vink10}, could explain the steep drop trend in the observed rotation of BSGs below $\sim 22$ kK -- the so-called ``bi-stability breaking''. However, \citep{Keszthelyi17} argued that young O stars could alternatively lose rotational speed already before reaching the BSJR due to internal structural changes, thus questioning the need for a mass-loss induced breaking mechanism. Given that our results point in the direction of the \citet{Krticka21} recipe, we would expect an increase in the loss of angular momentum, which is qualitatively also in line with the obtained low rotational velocities for our targets ($\sim 30$ km\,s$^{-1}$). However, whether such an increase in $\dot{M}$ provides a quantitatively sufficient rotational breaking mechanism is yet to be tested by evolutionary models.

\section{Evolutionary context}
\label{sec:evol_discuss}

\subsection{Possible mass-discrepancy in BSGs}
\label{sec:mass-discrepancy}

\begin{figure}
  \resizebox{\hsize}{!}{\includegraphics{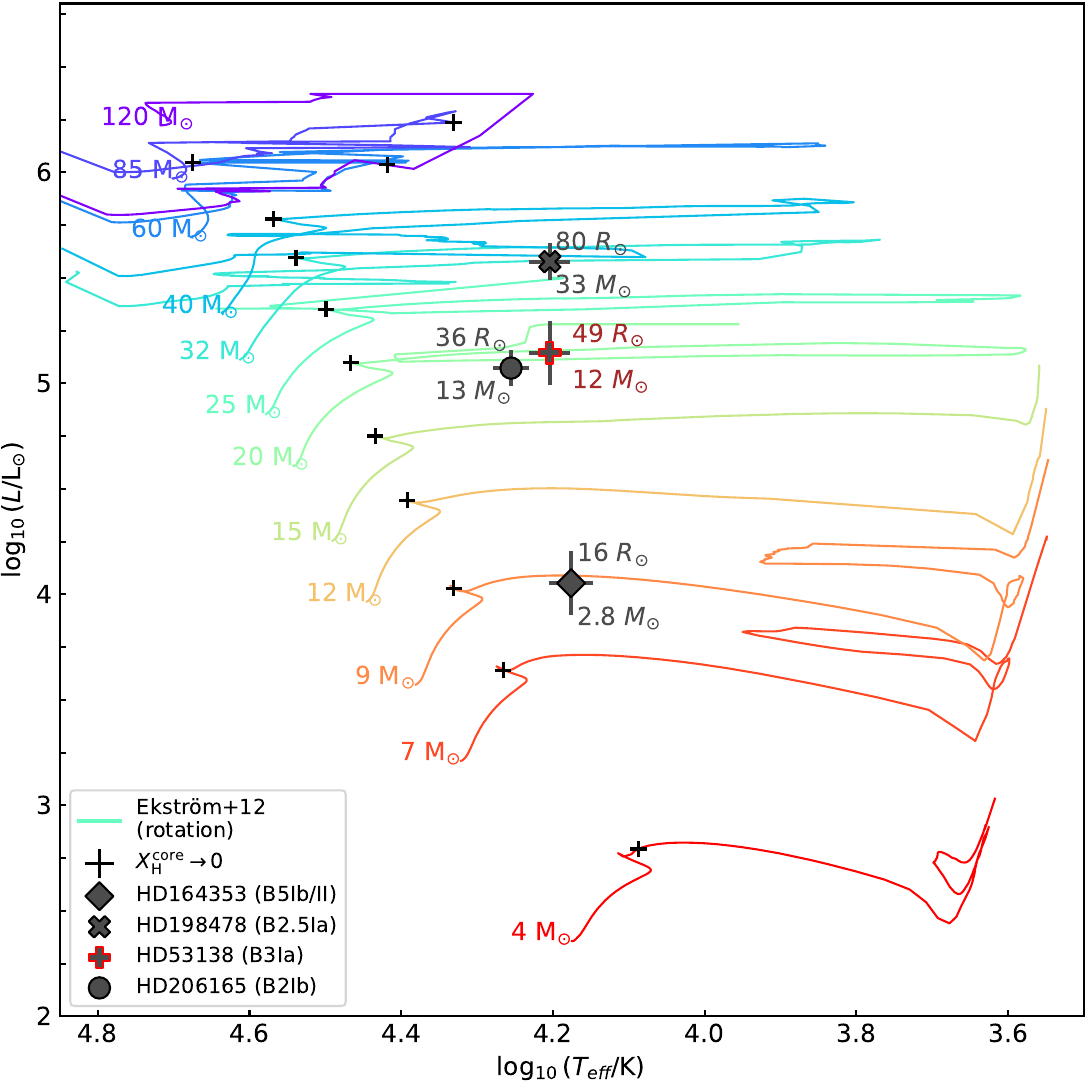}}
  \caption{Position of our sample BSGs on the HR Diagram in comparison with evolutionary tracks ($Z_\text{ZAMS} = 0.014$, with rotation) from \cite{Ekstrom12}. The numbers at the beginning of the tracks represent the initial stellar mass and the black thin crosses indicate where the star leaves the main sequence ($X^{\mathrm{core}}_{\mathrm{H}} \rightarrow$  0). The cross with the red stroke marks the position of HD53138 (B3Ia).}
  \label{fig:HRD}
\end{figure}

From the obtained stellar (photospheric) parameters, we can place our sample BSGs on the Hertzsprung-Russell diagram (HRD), presented in Fig.\,\ref{fig:HRD} and compared with evolutionary tracks from \cite{Ekstrom12} at solar metallicities (assuming an initial rotation of $\varv/\varv_\text{crit} = 0.4$). 

\begin{figure}
  \resizebox{\hsize}{!}{\includegraphics{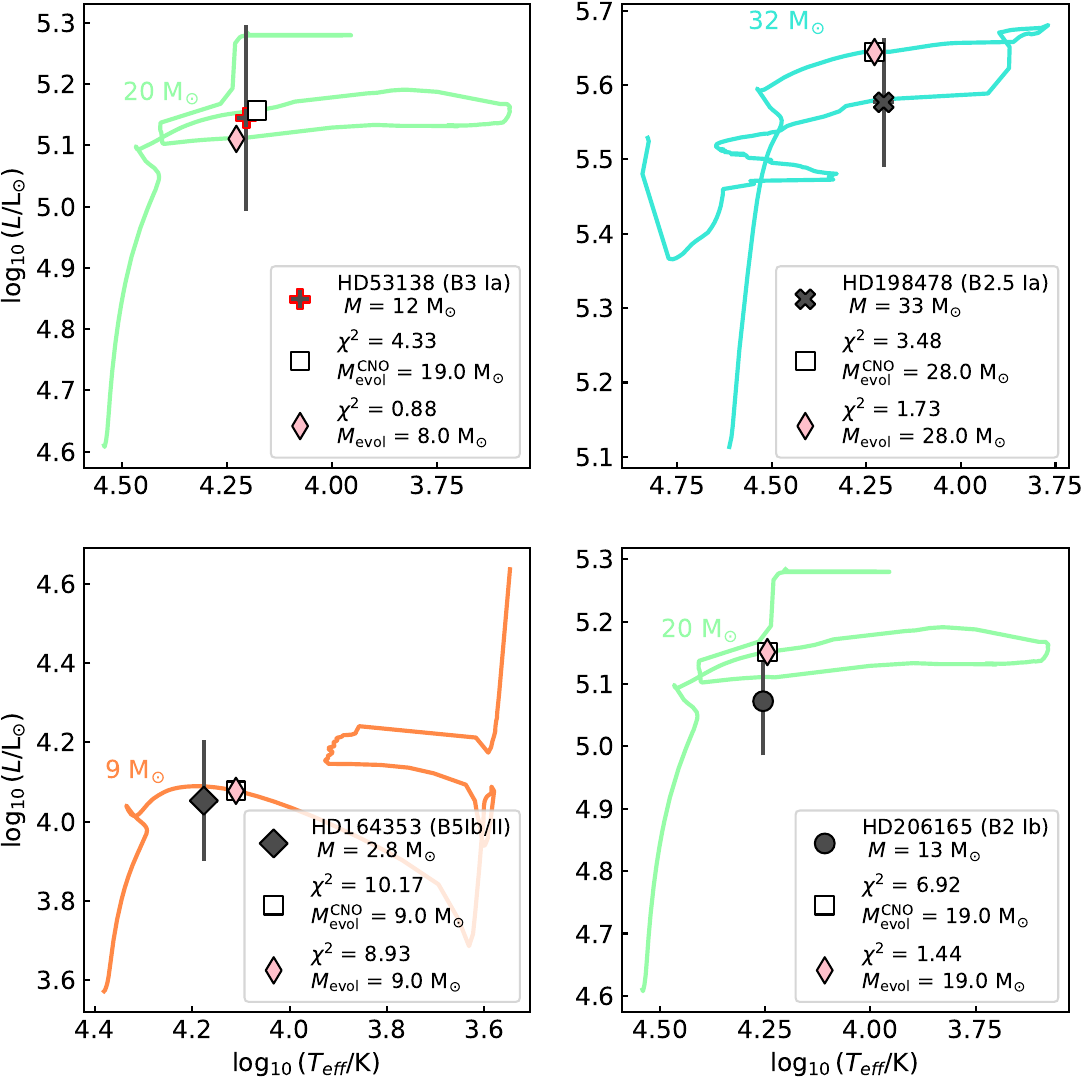}}
  \caption{Individual comparison of each BSG in the HRD with each of the best-fitted points (through an $\chi^2$ minimization) at the evolutionary tracks of \citet{Ekstrom12}. The pink diamonds indicate the best model considering $T_{\mathrm{eff}}$, $L$, $\log g$, while the white squares indicate the same but with the CNO abundances taken into account. For most stars, except HD53138, the result is exactly the same.}
  \label{fig:HRD_chi2}
\end{figure}

Both the position in the HRD and their surface CNO abundances place our sample BSGs as evolved  stars, displaced from the main sequence. For such objects, we would expect that their surfaces display (significant) chemical alteration due to internal mixing processes \citep[e.g.][]{Ekstrom12}. For BSGs that evolve directly off the main sequence towards the red supergiant (RSG) stage\footnote{Typically, in an evolutionary context, a massive star is considered to be a RSG if its $T_\mathrm{eff} \lesssim 5000$ K \citep[e.g.][]{Meynet15}.}, it is expected that the mixing occurs due to overshooting and rotation-driven large-scale circulation \citep[e.g., meridional circulation][]{Sweet50,Maeder09}.
Alternatively, stars can also become BSGs after having experienced a RSG stage. For these objects, one would expect higher surface chemical alteration due to internal convection at the RSG phase, and a reduced envelope due to the mass loss experienced in that extreme stage. This implies that post-RSG BSGs have a higher $L/M$ ratio, and, consequently, different pulsational properties compared to pre-RSG BSGs as well \citep[cf.][]{Saio13,Georgy14,Georgy21}.

In this paper, we perform an exploratory analysis of their status by comparing them with recent evolutionary tracks \citep{Ekstrom12}. We determine the best-matching evolutionary models for our BSGs by performing a $\chi^2$ fitting procedure taking into account different stellar parameters \citep[see, e.g., Sect.\,2 of][for an application using the Bonn tracks]{Schneider14}.
The absolute $\chi^2$ values as such do not yield any judgment about our derived physical quantities, but provide us with a systematic approach to select the best-describing structure model from a given set of evolutionary tracks.

Initially, we considered $L$, $T_\mathrm{eff}$, and $\log g$ for finding the best evolutionary track, as these are the main stellar properties and can directly be obtained via quantitative spectroscopy, alongside knowledge about the distance. The results are shown in Fig.\,\ref{fig:HRD_chi2}.
From the $\chi^2$-fitting, the estimated evolutionary masses are higher than the spectroscopic masses for HD164353 and HD206165, albeit still within the mass uncertainty range for the latter. HD198478 (B2.5Ia) reasonably agrees with the closest available track ($M_{\text{ZAMS}} = 32\,\mathrm{M_\odot}$) and HD53138 is best fitted by a $8\,\mathrm{M_\odot}$ post-RSG model ($M_{\text{ZAMS}} = 20\,\mathrm{M_\odot}$). The mass discrepancy is most pronounced in the case of HD164353, in which the evolutionary mass is about three times the spectroscopic determination. However, our derived $\chi^2$ value for this object is also the largest one among our sample, thus indicating that HD164353 might not be represented by any of the evolutionary tracks considered so far.

Given the properties of BSGs and their position in the HRD, it is not obvious whether individual objects are pre- or post-RSG stars. From the $\chi^2$ minimization procedure, considering $L$, $T_\mathrm{eff}$ and $\log g$, we find all but one object (HD53138) to be pre-RSGs.

One way proposed to discriminate their evolutionary status is via analyzing the surface CNO abundances \citep{Hunter09,Saio13}\footnote{Some physical properties such as the rotational velocities \citep{Zorec09} and modeling choices such as the adoption of the Ledoux or Schwartzschild criterion might affect the abundances as well \citep{Georgy21}}. If we also take our derived surface CNO abundances at face value and account for them in the $\chi^2$ fit (cf.\ the squares in Fig.\,\ref{fig:HRD_chi2}), even HD53138 is considered to be a pre-RSG object. In that case, the spectroscopic mass ($13$ M$_{\odot}$) would be lower than the evolutionary model's value ($19$ M$_{\odot}$).

Concerning the evolutionary status, when we compare our results with previous investigations, we find that for HD53138, \citet{Martin18} \citep[citing ][]{Zorec09} argued that there is no evidence that this BSG is a post-RSG object. For HD198478, \cite{Kraus15} showed evidence that this object would be categorized as a post-RSG star due to its pulsational behavior ($\alpha$\,Cyg variable) and the presence of a bow-shock feature in WISE images (4 $\mathrm{\mu}$m < $\lambda$ < 24 $\mathrm{\mu}$m), expected for stars with a previous phase of intense winds \citep[e.g.,][]{Gordon19b}.

If we assume that our targets are pre-RSG objects (as indicated by the $\chi^2$ fitting procedure accounting for the CNO abundance) our findings would suggest that three of our four cool BSGs might be affected by the so-called ``mass discrepancy problem'' \citep[e.g.,][]{Herrero92,Mokiem07,Searle08,Cantiello09}, which is commonly discussed for OB stars, where systematic differences are obtained between the masses inferred from spectroscopy and those derived from their luminosity via evolutionary models.

In Table~\ref{tab:probabs}, we quantify the possibility of a mass discrepancy for our sample by computing the probability of deriving lower spectroscopic masses with respect to the masses obtained from the $\chi^2$ results. Thereby, the uncertainties in $\log L$, $T_\mathrm{eff}$ and $\log g$ are taken into account (see Appendix Sect.~\ref{app:statistics}). We integrated the spectroscopic mass distribution for each star up to the value derived from the evolutionary tracks. For two cases, HD198478 and the putative post-RSG HD53138, the obtained evolutionary mass is close to the peak of the distribution. For the other two objects, however, the bulk of the distribution is lower than the mass from the evolutionary tracks (see the $P_\mathrm{lower}$ value in Table \ref{tab:probabs}). Among these, HD206165 has a reliable Gaia DR3 distance and thus could be considered as a more robust example of a mass-discrepant BSG.

\begin{table}
\caption{\label{tab:probabs}Probability of deriving a lower spectroscopic mass by chance for each star.}
\centering
\begin{tabular}{ccccccc}
\hline
\hline
Star     & $P_\mathrm{lower}$ & $M_\mathrm{spec}$ & $M_\mathrm{evol}$ & $M_\mathrm{evol}^\mathrm{CNO}$ \rule[-1.5mm]{0mm}{4.8mm} \\
\hline
HD206165 & 0.86   & 13    & 19       & 19               \\
HD198478 & 0.34   & 33    & 28       & 28           \\
HD53138  & 0.23   & 12    & 8        &  19           \\
HD164353 & 0.99   & 2.8   & 9        &  9            \\
\hline
\end{tabular}
\end{table}

Among our sample, HD164353 exhibits the most extreme discrepancy with a spectroscopic mass of less than 3 $\mathrm{M_\odot}$, while the corresponding evolutionary track describes a star with $M_\mathrm{ZAMS} = 9\,\mathrm{M_\odot}$. From Table~\ref{tab:probabs}, the obtained value of $P_\mathrm{lower}$ indicates that it is extremely unlikely that we obtain this value by chance. From a single-star evolution perspective, a star born with $\sim 9\,\mathrm{M_\odot}$ would not be able to lose about 2/3 of its mass in an instrinsic way.

Errors in $\log g$ and $L$ could be possible explanations. With higher $\log g$, we could obtain higher masses. For example, with $\log g = 2.75$, as found by \cite{Searle08}, we would obtain 7.4\,M$_\mathrm{\odot}$, which is much closer to the evolutionary track.

    \begin{figure}
      \resizebox{\hsize}{!}{\includegraphics{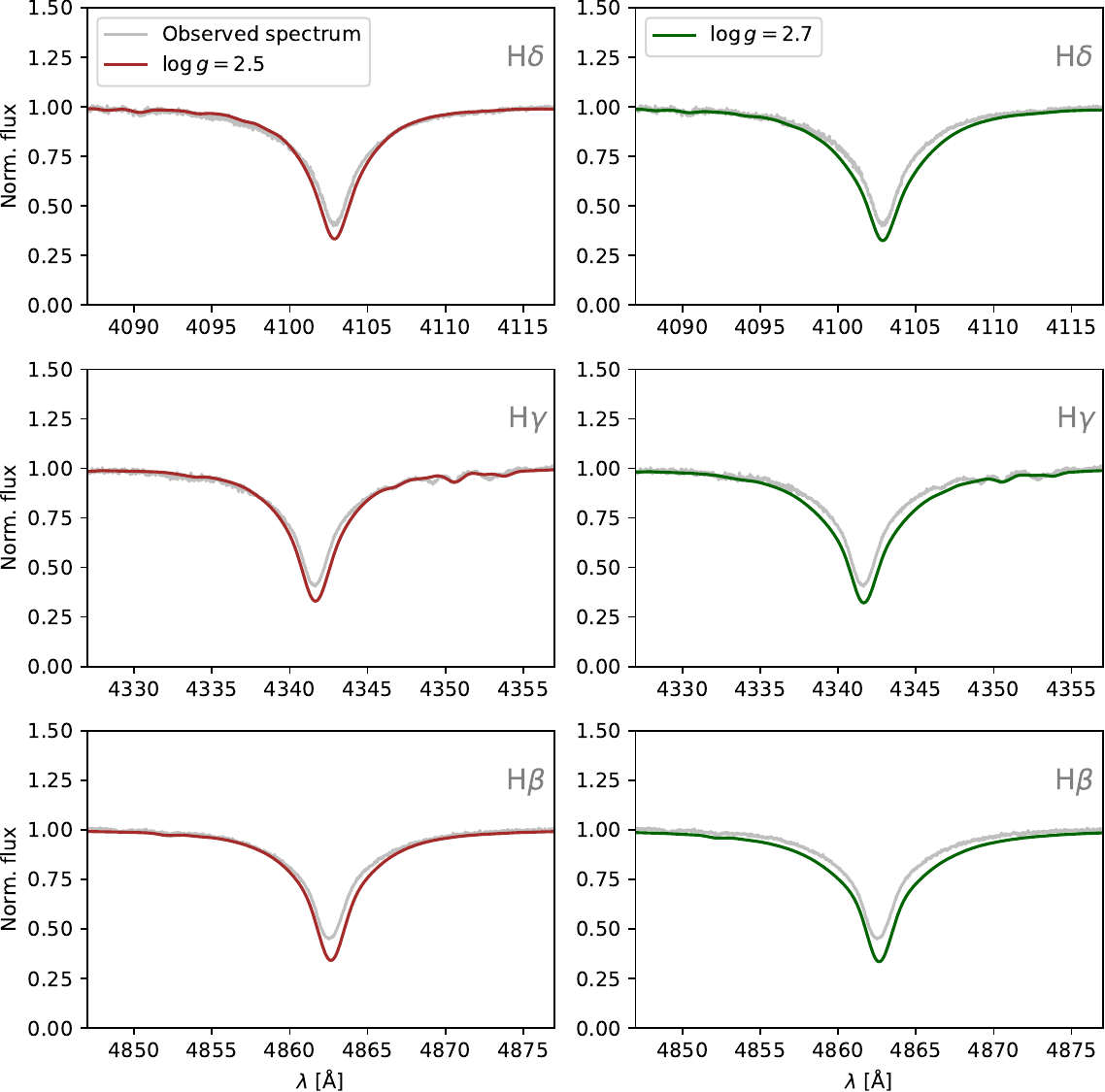}}
      \caption{Comparison between models with different $\log g$ for HD164353. The left panels show the Balmer lines for the model with $\log g = 2.5$ (brown lines), while the right panels depict the model with $\log g = 2.7$ (green line).}
      \label{fig:logg_comp_snake}
    \end{figure}

However, in our study, such a high value would yield significantly too wide Balmer line wings. This is illustrated in Fig.\,\ref{fig:logg_comp_snake}, where we compare different models with $\log g$ of $2.5$ and $2.7$, respectively. Alternatively, one could see this as a possible issue with the distance and thus the derived luminosity. If we use the Gaia eDR3 distance, as recently done by \citet{Wessmeyer22}, we would obtain a higher luminosity and, thus, a higher mass ($M = 13$ M$_\mathrm{\odot}$), which would yield properties more similar to other sample BSGs. However, as the Gaia parallax for this object is indicated as unreliable, we chose the Hipparcos parallax as discussed in Sect.\,\ref{sec:observation}.  

Besides all the obtained statistics, there are important caveats to consider when trying to infer the evolutionary status of an object from comparing spectroscopic values to evolutionary tracks. First, we are limited to a discrete set of tracks with predetermined initial masses and initial rotation. 
Moreover, the masses and surface abundances of post-RSG stars are particularly uncertain as their properties depend on all the physics and assumptions included in the former evolutionary stages \citep[e.g.,][]{Martins13}. For example, the lifetimes of massive stars in the BSG regime can depend significantly on the assumptions related to the internal mixing, such as the overshooting parameter \citep[e.g.,][]{Schootemeijer19,Higgins23}.
Moreover, the mass-loss rates in the RSG stage are still quite uncertain \citep[see, e.g, the recent review by][]{Decin21}. Some recent empirical studies yield lower values for $\dot{M}$ \citep[e.g.,][]{Beasor21}. However, if RSG mass-loss rates would be lower during the whole RSG lifetime, the resulting synthetic populations might be at odds with observed populations \citep[e.g.,][]{Massey23}. Consequently, our CNO abundances are of limited diagnostic value as they could potentially also be reached in a post-RSG stage.

Investigation into whether the ``mass-discrepancy problem'' is manifested in cool BSGs was previously undertaken \citep[e.g.,][]{Searle08, Trundle-Lennon05}. Nonetheless, the comparisons between the spectroscopic results and evolutionary models were done ``by eye.'' From our analysis, given the small size of the sample, it is not possible to affirm as it does for other categories of OB stars. However, we present an approach that allow us to deliberate the existence of the problem in a mode systematic way. Thus, performing such an analysis on a larger sample would yield more concrete answers to this question.

\subsection{Investigating multiplicity}
\label{sec:multiplicity}

In general, BSGs are difficult to understand in the standard picture of stellar evolution because they are observed in a region of the HRD where massive stars spend very little time (thermal-timescale evolution through the Hertzsprung gap) assuming ``typical'' mixing settings. For example, the best-fitting single-star models found here spend $\lesssim 10^4$ years in that region.
Stellar mergers in binary stars can lead to the formation of stars that spend their entire core-helium burning lifetimes as BSGs \citep[e.g.,][]{Hellings83, Braun95, Podsiadlowski89, Podsiadlowski90, Claeys11, Vanbeveren13, Justham14}. 

\begin{figure}
  \resizebox{\hsize}{!}{\includegraphics{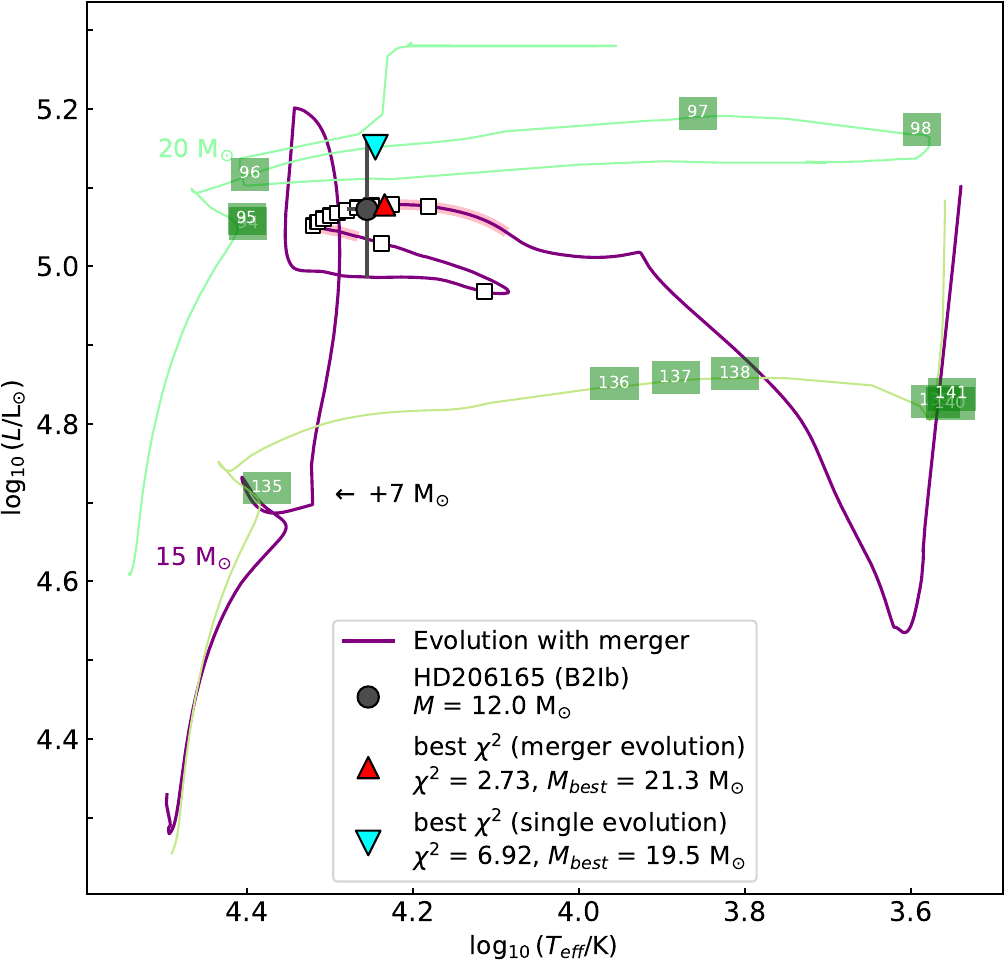}}
  \caption{Comparison between single evolutionary tracks \citet{Ekstrom12} and a MESA model with a merger (purple broader line) for HD206165. The squares with numbers are the ages in units of $10^5$ years for the \citet{Ekstrom12} tracks while the white squares show the same time difference for the merged model. The shaded region in pink shows the core-He burning phase. The arrow indicates the point of the $15$~M$_\odot$ evolutionary track where the merging (mass addition of 7~M$_\odot$) happened.}
  \label{fig:HRD_merger}
\end{figure}

To understand whether merging can explain the observed stellar properties of the BSGs studied here, we mimic the merger process by rapid accretion of mass onto stars that just finished core hydrogen burning. In that procedure, we do not account for alteration in the surface chemical composition in the merging process. We employed MESA revision 10398 \citep{Paxton11, Paxton13, Paxton15, Paxton18, Paxton19} and use the same model setup as in \citet{Schneider21}. The merged stars can simultaneously reproduce the observed luminosities, effective temperatures and surface gravities while burning helium in their cores (see e.g., Fig.~\ref{fig:HRD_merger}).

As with the single-star models, we tend to find masses higher than the spectroscopic determinations, especially for HD164353, namely, the $L/M$-ratio of the relaxed merger model is not much different from the $L/M$ ratio of a single-star evolutionary model with a higher initial mass. Therefore, it does not resolve the mass discrepancy problem. However, the merger models spend $\sim 10^5$ yr in the location of the HRD where the BSGs are observed, making a merger solution for them much more likely compared to a single-star scenario. 
This idea can be further investigated by, for example, searching for possible, close binary companions, and having more detailed merger simulations that trace  the alterations in surface chemical abundances for comparison with the measured composition.
This idea can be further investigated by, for example, searching for possible, close binary companions, and having more detailed merger simulations that trace the alterations in surface chemical abundances, for comparison with the measured composition.
Although it does not provide a clear solution to the mass discrepancies, this result provides evidence that many BSGs could be products of stellar mergers.

Alternatively, the discrepant masses could be the result of stellar stripping. There are indications that some partially stripped stars can appear as blue supergiants.
As discussed for example in \citet{Klencki22}, such objects would display a lower spectroscopic mass (relative to the ``evolutionary mass''), CO depletion plus N enrichment, and slow rotation. All of these characteristics are present in most of our sample stars, especially in HD164353. However, their results were obtained at LMC metallicity and the scenario did not occur at solar metallicity. Moreover, much lower $\log g$ values than our determinations are expected.
On the other hand, \citet{Irrgang22} analyzed the B2.5 subgiant $\gamma$\,Col and identified it as a stripped, pulsating core of a previously more massive star. They found a strong nitrogen enrichment of $\sim 10$ times the solar value. While our sample stars do show N enrichment, our factors of up to 4 are significantly below this value.

For our targets, there is no real evidence for existing companions so far. Of course, the large luminosities of BSG-like stars can easily outshine darker companions, but for systems that interacted in the past one would expect to see radial velocity shifts when comparing spectra from different epochs, as for example in the high-mass X-ray binary (HMXB) QV Nor (B0I+NS), which has a mass ratio of $\sim$15 and a clearly detected orbital motion of the donor star \citep{Reynolds92}. In the case of such compact companions, the detection of a high X-ray luminosity, caused by the accretion of (wind) material on the compact object, would be a clear indication. However, as we discuss in Sect.\,\ref{sec:wind_analysis}, our sample does not contain objects that are overluminous in the X-ray regime. Moreover, wind-fed HMXBs typically have B0-B1 type donors \citep{MartinezNunez17}.
From the current binary evolutionary scenarios and their predictions, it thus seems unlikely that our sample stars are part of a binary system with a ``recent'' interaction.

In summary, the derived properties of our BSGs, except for the luminosity and mass of HD164353, are in the range of properties found in other BSG studies \citep[cf.][]{Searle08, Fraser10, Haucke18}. Our obtained mass discrepancy is relatively common, which leaves the puzzling question of whether the observed population of BSGs is dominated by immediate post-MS stars or consists of a significant number of more evolved objects such as post-RSG stars, stellar mergers, or partially stripped post-interaction binary products. A more extensive and coherent sample study, such as provided by ULLYSES \citep{Roman-Duval2020} and its accompanying X-Shooter program \citep[XShootU,][]{Vink+2023}, as well as (if possible) more precise mass measurements will be key to answering this question.

\section{Conclusions}
\label{sec:conclusion}

In this study, we performed an in-depth spectral analysis of four Galactic cool BSGs with CMFGEN and PoWR, covering the UV and optical regime. For the first time in this regime, we simultaneously reproduced the observed optical spectrum together with major wind-affected UV profiles. We obtained the photospheric and wind properties of our sample stars as well as constraints on their clumping and X-ray properties. From the immediate spectral analysis, we drew the following conclusions:

\begin{itemize}
    \item By including clumping, and especially X-rays, in the atmosphere models of cool BSGs, we can fit key UV wind lines and generally reproduce the observed superionization. This represents a major improvement in spectral modeling of later type BSGs. 
    
    \item We find lower relative X-ray luminosities than the ``canonical'' relation of $\log (L_\mathrm{x}/L) = -7.0$. Assuming a shock temperature compatible with the typical wind velocities of our sample stars ($T_\mathrm{x} = 0.5$), we obtain $\log (L_\mathrm{x}/L)$ in the range of $-8.3$ to $-7.3$. 

    In some spectra, it is unclear whether \ion{N}{V} is actually present. Disregarding \ion{N}{V}, we find
    even lower limits of $\log (L_\mathrm{x}/L) \sim 13.3$ and shock temperatures of only $0.1\,$MK. This happens independently of the adopted clumping values. 
    Revisiting cooler BSGs with new X-ray observations 
    will be necessary to put more robust constraints on the wind-intrinsic X-ray emission of these stars.
    
    \item Our results produce further evidence that the winds of BSGs are smoother than the winds of their hot counterparts as recent hydrodynamical simulations predict \citep{Driessen19}. In the microclumping formalism, our targets are described by volume filling factors of $0.5$ to $1.0$. Additionally, we found evidence for the presence of optically thick clumping, which improves the H$\alpha$ fit when accounted for in the spectral modeling. 
\end{itemize}

From our best-fit models, we investigated the consequences of the derived stellar and wind parameters on their evolutionary status:

\begin{itemize}
    \item We find stellar properties compatible with evolved stars, namely altered CNO abundances and a clear displacement from the main sequence in the HRD.
    Our results are in line with previous literature providing properties of cool BSGs. 
    
    \item For two of our four targets, our derived spectroscopic masses are lower than those expected from immediate post-MS stars assuming single-star evolution (including rotation). Our systematic analysis indicates tentative evidence that the so-called ``mass discrepancy problem'' that is known for many OB stars might also apply to at least some BSGs. A more in-depth look into this problem with a larger sample is necessary to get a more general overview of the situation in the BSG regime.
        
    \item Our results indicate that at least a sub-sample of cool BSGs might not stem from a single-star evolution. Given the intrinsically high binary fraction among massive stars, it is likely that some BSGs are products of binary interaction (e.g., stripping or merger). We find possible merger scenarios to explain some of our objects, but the corresponding evolution models do not resolve the mass discrepancy problem.
\end{itemize}

Besides the individual status of each object, we can also draw more general conclusions about massive stars and their evolution from the derived wind parameters:

\begin{itemize}
    \item Our derived mass-loss rates are systematically lower than values from previous studies, albeit usually within  the uncertainty range. We attribute this to the effect of the included X-rays on the UV P\,Cygni profiles used for mass-loss diagnostics.
    
    \item When comparing our findings to different mass-loss recipes, we find that our cool BSGs are significantly lower than predicted by \citet{Vink01}, but generaly align remarkably well with the predictions by \citet{Krticka21}. Including early-B and O supergiants for comparison, we observe the same agreement. For these earlier-type stars, there is a good agreement between the predictions by \citet{Krticka21} and \citet{Bjoerklund22}. This, however, does not happen for the late BSGs, where the \citeauthor{Bjoerklund22} rates seem to be too low. 
    
    \item In contrast to the \citet{Bjoerklund22} predictions, our findings provide strong evidence that an increase in the mass-loss rates does occur when a supergiant crosses the bi-stability jump region during its evolution. The increase, however, seems to be less pronounced than predicted by \citet{Vink99,Vink00,Vink01}, but rather on the level predicted by \citet{Krticka21}. Interestingly, the characteristic temperature of the switch seems to align more with the predictions of \citet{Vink99} than the \citet{Krticka21} recipe. A detailed wind-driving investigation on the bi-stability region will be necessary to draw more robust conclusions about where and under which circumstances the mass-loss rate increases in this regime.

\end{itemize}

Overall, our study reveals the importance of accounting for X-rays in the study of cooler B supergiants, despite a so far quite limited observational constraints. Moreover, our findings underline that the puzzling status of B supergiants is far from being solved and a larger sample study with multi-wavelength observations is required. Additional information from variability studies and asteroseismology as well as theoretical and numerical studies -- such as dynamically-consistent atmospheres, 3D structure modelling, or interacting binary evolution -- will be required to fully determine the nature of these crucial objects at the crossroads of massive star evolution.

\begin{acknowledgements}
      The author would like to thank the referee who provided insightful and constructive comments that helped to improve this manuscript. MBP, AACS, and VR are supported by the German
      \emph{Deut\-sche For\-schungs\-ge\-mein\-schaft, DFG\/} in the form of an Emmy Noether Research Group -- Project-ID 445674056 (SA4064/1-1, PI Sander).
      MPB, AACS, and VR further acknowledge support from the Federal Ministry of Education and Research (BMBF) and the Baden-Württemberg Ministry of Science as part of the Excellence Strategy of the German Federal and State Governments.
      FRNS received funding from the European Research Council (ERC) under the European Union’s Horizon 2020 research and innovation programme (Grant agreement No.\ 945806) and is supported by the Deutsche Forschungsgemeinschaft (DFG, German Research Foundation) under Germany’s Excellence Strategy EXC 2181/1-390900948 (the Heidelberg STRUCTURES Excellence Cluster). F.N. acknowledges grant PID2019-105552RB-C41 funded by the Spanish  MCIN/AEI/ 10.13039/501100011033.
\end{acknowledgements}

\bibliographystyle{aa}
\bibliography{biblio.bib}

\appendix

\section{Quantifying uncertainties in mass and radius}
\label{app:statistics}

\begin{figure}
  \resizebox{\hsize}{!}{\includegraphics{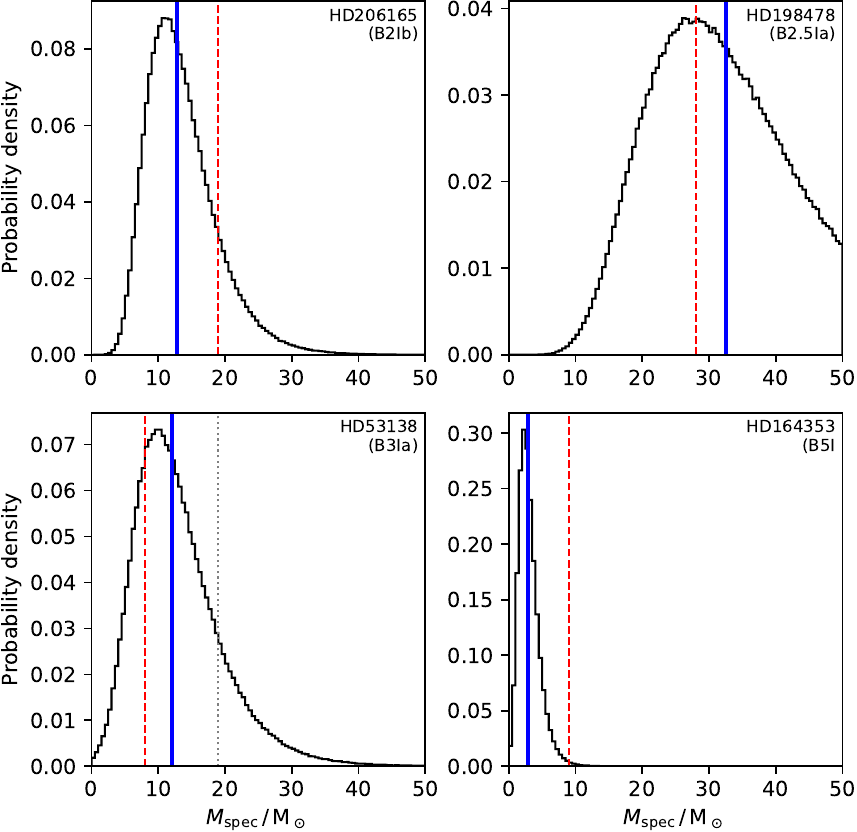}}
  \caption{Histograms depicting the probability distribution of the masses of each BSG in the sample. The blue-filled and red-dashed lines indicate, respectively, the spectroscopic and evolutionary masses obtained. The grey dotted line for HD53138 indicates the evolutionary mass obtained accounting for the CNO abundance.}
  \label{fig:histograms}
\end{figure}

In this study, we quantify the errors in $M$ and $R$ by considering that the errors in luminosity ($L$), effective temperature ($T_\mathrm{eff}$) and surface gravity ($\log g$) follow a Gaussian distribution around their nominal values for each star. Based on this, we then compute the mass distribution for each BSG. The mean value of each distribution corresponds to the nominal value of $M$ (or $R$) and the standard deviation is adopted as the corresponding uncertainty. The results of this procedure are illustrated for $M$ in Fig.~\ref{fig:histograms}, where we show the mass distribution for each star together with the predicted evolutionary mass from the $\chi^2$-fit (see Sect.~\ref{sec:evol_discuss}). 

\section{UV wind lines of the models with different clumping and X-rays combinations.}
  \label{app:UV_fits}
  
\begin{figure*}
\centering
   \includegraphics[width=17cm]{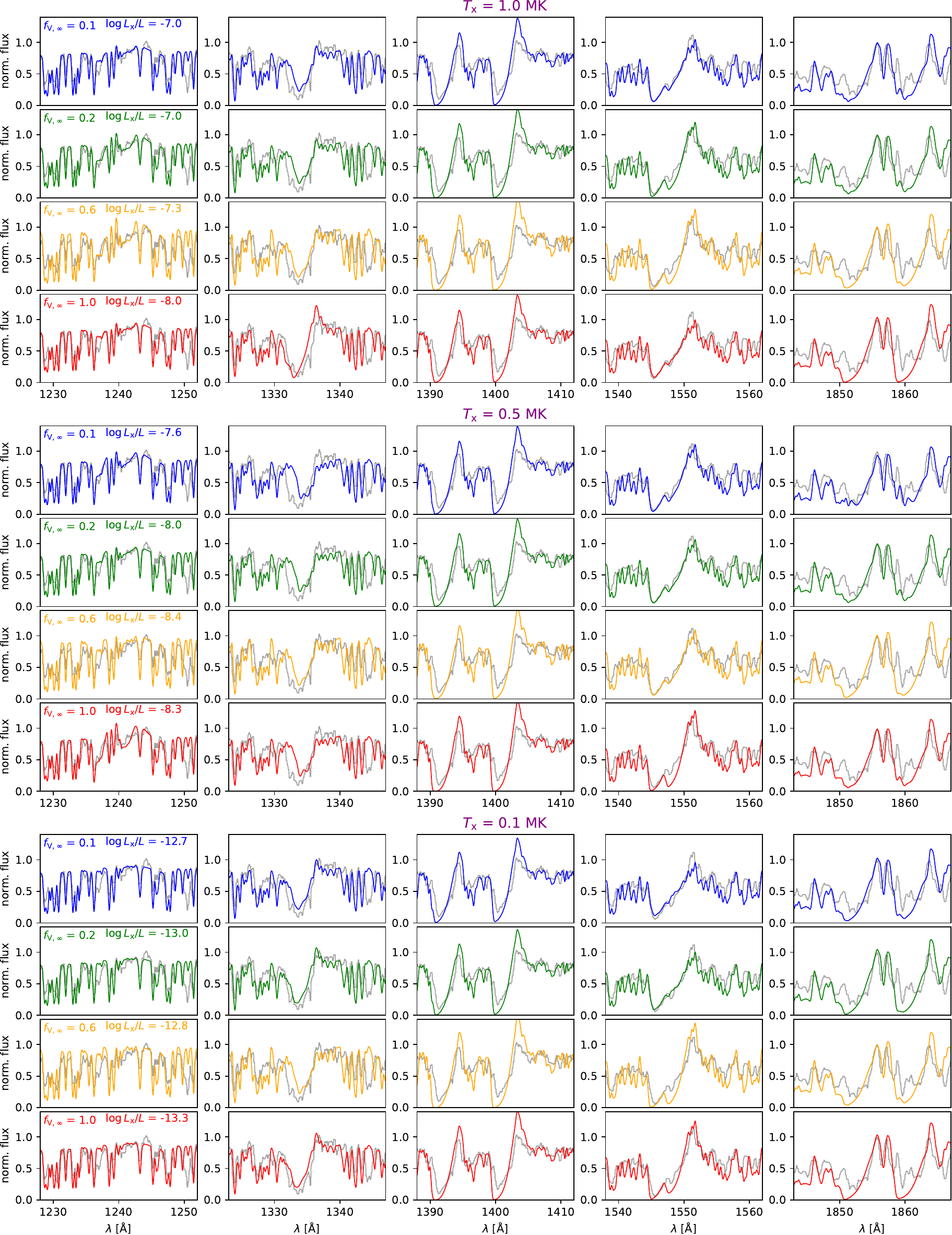}
     \caption{Best-fitting models for HD53138 in the UV region, focusing on the main P-Cygni profiles, namely \ion{N}{V} $\lambda$1238-42, \ion{C}{II} $\lambda$1335-36, \ion{Si}{IV} $\lambda$1394-1403, \ion{C}{IV} $\lambda$1548-50 and \ion{Al}{III} $\lambda$1855-63, depicted in each column. Each group of four rows has the same shock temperature, in which each row has a model with an assigned clumping and X-ray luminosity.}
     \label{fig:general_UV_fitting}
\end{figure*}

\begin{figure*}
\centering
   \includegraphics[width=17cm]{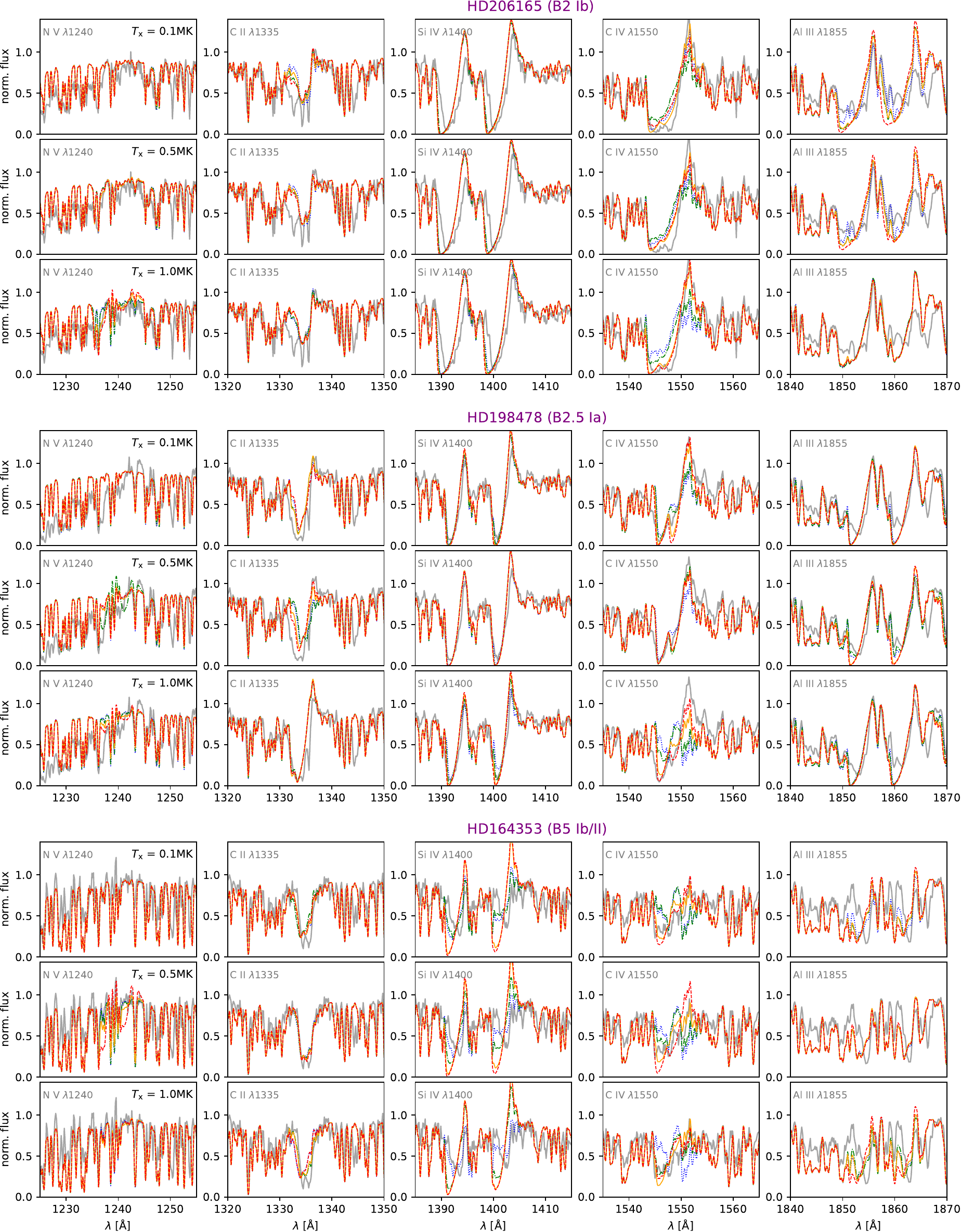}
     \caption{Best-fitting models for the main UV lines of HD206165, HD198478 and HD164353. Each row of plots shows the lines of the same model with a given shock temperature $T_\mathrm{x}$. The different colors/line styles represent the different adopted values of clumping, namely: red-dashed stands for $f_\infty = 1.0$, orange/full for $f_\infty = 0.5$, green-dash-dotted for $f_\infty = 0.2$ and blue-dotted for $f_\infty = 0.1$. The associated $\log (L_\mathrm{x}/L)$ and $\dot{M}$ for each clumping and shock temperature value are displayed in Tables \ref{tab:LXs_allstars} and \ref{tab:wind_prop}.}
     \label{fig:general_UV_fitting_other_BSGs}
\end{figure*}

In this appendix, we display the models for each clumping and X-rays settings in more detail for HD53138 and for the other sample stars (see Figs~\ref{fig:general_UV_fitting} and \ref{fig:general_UV_fitting_other_BSGs} respectively).

\section{Additional information and comments about the sample BSGs}
\label{app:info_other_BSGs}
\subsection{HD206165 (B2Ib)}

We note that 9 Cep is a star that has not been studied in great detail previously. Given its spectral type, it was assigned a temperature of $\sim 20$ kK, very close to the bi-stability jump, by \citet{McErlean99}. Its spectrum was modeled by \cite{Searle08} and \cite{Markova08}, using  CMFGEN and FASTWIND respectively. \citeauthor{Searle08} found a temperature and surface gravity similar to our values while \citeauthor{Markova08} found a similar $\log g$ but higher temperature ($19.6\,$kK).

The Combined General Catalogue of Variable Stars \citep[CGVS][]{Samus04} shows that the star has a magnitude variation of $\sim 0.1$ and classified it as an $\alpha$-Cygni variable (but with no evidence of periodic pulsation). However, we did not notice any meaningful alteration in the spectral lines between different available data (2007 and 2011).

For the wind terminal velocities, both authors following \cite{Howarth97} used $\varv_\infty = 640\,\mathrm{km}\,\mathrm{s}^{-1}$. However, we could only obtain a good fit for the UV P-Cygni lines -- except for \ion{C}{II}, which is even broader -- when applying $\varv_\infty = 900\,\mathrm{km}\,\mathrm{s}^{-1}$. 

\subsection{HD198478 (B2.5Ia)}

HD198478 is a BSG that has been analyzed in greater detail in the past decades by different authors using multiple techniques, namely: \cite{Crowther06} and \cite{Searle08} who used CMFGEN to model its optical+UV spectrum, \cite{Kraus15} who employed FASTWIND to model tens of optical spectra collected in a five-year campaign, and \cite{Gordon19} who measured some of its properties using interferometry from the CHARA array. This makes HD198478 one of the most well-studied cooler BSGs so far.

Despite the multiple analyzes performed for this, there is quite a broad discrepancy between these different works regarding the obtained properties for HD198478. For instance, \cite{Crowther06} (similarly to our work) found $T_\mathrm{eff} = 16.5\,$kK and $\log g = 2.15$ using CMFGEN, while \cite{Kraus15} used FASTWIND to determine $T_\mathrm{eff} \approx 19.0\,$kK and $\log g \approx 2.4$. \cite{Markova08}, also  using FASTWIND, found an intermediate value of $17.0$\,kK. 

Regarding the wind velocity, the situation is equally discrepant with terminal velocities found in the range between $200$ and $250\,\mathrm{km}\,\mathrm{s}^{-1}$ \citep{Markova08,Kraus15} and $550\mathrm{km}\,\mathrm{s}^{-1}$ \citep{Searle08}. In this work, we derived $\varv_\infty = 570\mathrm{km}\,\mathrm{s}^{-1}$. 

The main physical parameters obtained for HD198478 are notably close to those of HD53138 (B3Ia) -- which is not totally surprising given that they have a very similar spectral type and their optical spectra are almost identical.

\subsection{HD53138 (B3Ia)}

24-CMa or $o^2$-CMa is a relatively well-studied BSG, hence, it is our choice for a more in-depth analysis. Its atmosphere was modeled by several authors using CMFGEN and FASTWIND \citep[e.g.,][]{Crowther06,Lefever07,Searle08,Haucke18}.

As with many other BSGs, this star displays spectral variability in H$\alpha$ (as discussed in Sect.\,\ref{sec:observation}). Additionally, it also possesses weak photometric variability detected by TESS and Hipparcos \citep{Bowman22,Lefevre09}. From that variability and its relatively elevated macroturbulent velocity, \cite{Haucke18} also classified this star as an $\alpha$-Cygni variable.

HD53138 was further scrutinized by the LIFE project \citep{Martin18}, which seeks for magnetic fields in OB stars. However, no evidence of magnetic activity was detected.

\subsection{HD164353 (B5Ib/II)}

67 Oph is a BSG on the edge of the bright giant classification -- which is in line with its larger Balmer lines and higher inferred surface gravity. However, HD164353 has not been extensively studied in the literature. \cite{Zorec09} analyzed this star looking at broader features in the spectrum and reported a $T_\mathrm{eff}$ of $15.4\,$kK.

Currently, the only works which analyzed this star using a modern atmosphere code are \cite{Searle08} and \citet{Wessmeyer22}. \citeauthor{Searle08} derived a temperature ($T_\mathrm{eff} = 15.5$\,kK), luminosity ($\log (L/\mathrm{L_\odot}) = 4.3$) and radius (16 $\mathrm{R_\odot}$) that are similar to the values we obtained (see Table \ref{tab:phot_prop}). However, their $\log g$ is 0.3 dex higher than ours, which results in a stellar mass twice as large as ours. \citeauthor{Wessmeyer22} found nearly identical parameters to our study. However, they considered the Gaia EDR3 distance, which is twice the value from Hipparcos and we disregarded it due to its high RUWE value. Consequently, \citeauthor{Wessmeyer22} obtained a higher value for the luminosity and mass of HD164353.

\cite{Vaiana81} reported the detection of X-rays (>0.2 keV, using the Einstein X-ray telescope) with a $\log(L_\mathrm{X}/L) = -6.9$. Employing ROSAT, \cite{Berghoefer97} only obtained the same ratio, but only as an upper limit for the X-ray flux.

\section{Further details on the X-ray parameters}
\label{app:xray_params}

\subsection{Choice of X-ray onset velocity}

In CMFGEN, wind-intrinsic X-rays are implemented via an additional X-ray emissivity $\eta_\text{x}$ calculated via
\begin{equation}
    \label{eq:XR_cmfgen}
    \eta_\text{x}(r,\nu) \propto f_\text{x}^2 \, \, \,  e^{-\varv_\text{x}/\varv(r)} \, \, \, \frac{e^{-h\nu/k_\text{B} T_x}}{\sqrt{T_x}} \sum_j^{\mathrm{all \, ions}} z_j^2 g_\text{ff}(\nu,r) n_\text{e}(r) n_j(r),
\end{equation}
where $\varv_\text{x}$ is the characteristic velocity and $f_\text{x}$ is a filling factor, which translates the amount of shocks producing X-ray in the wind. $T_\text{x}$ is a characteristic temperature reached in the shocks, which controls the X-ray distribution in the frequency space. The quantities $n_e$ and $n_j$ are the density of electrons and ions respectively, $g_\text{ff}$ is the free-free Gaunt factor and $k_B$ and $z$ are, respectively, the Boltzmann constant and the charge of the ion $j$. 

Focusing on the space dependency (i.e., the $\exp[-\varv_\mathrm{x}/\varv(r)]$ term), $\varv_\text{x}$ determines the spatial distribution of the X-ray emissivity. The characteristic velocity describes the point where X-rays become meaningful in the wind, where the profile function would reach $\sim$0.37, namely, $\sim$37\% of the maximum reachable value.

\begin{figure}
  \resizebox{\hsize}{!}{\includegraphics{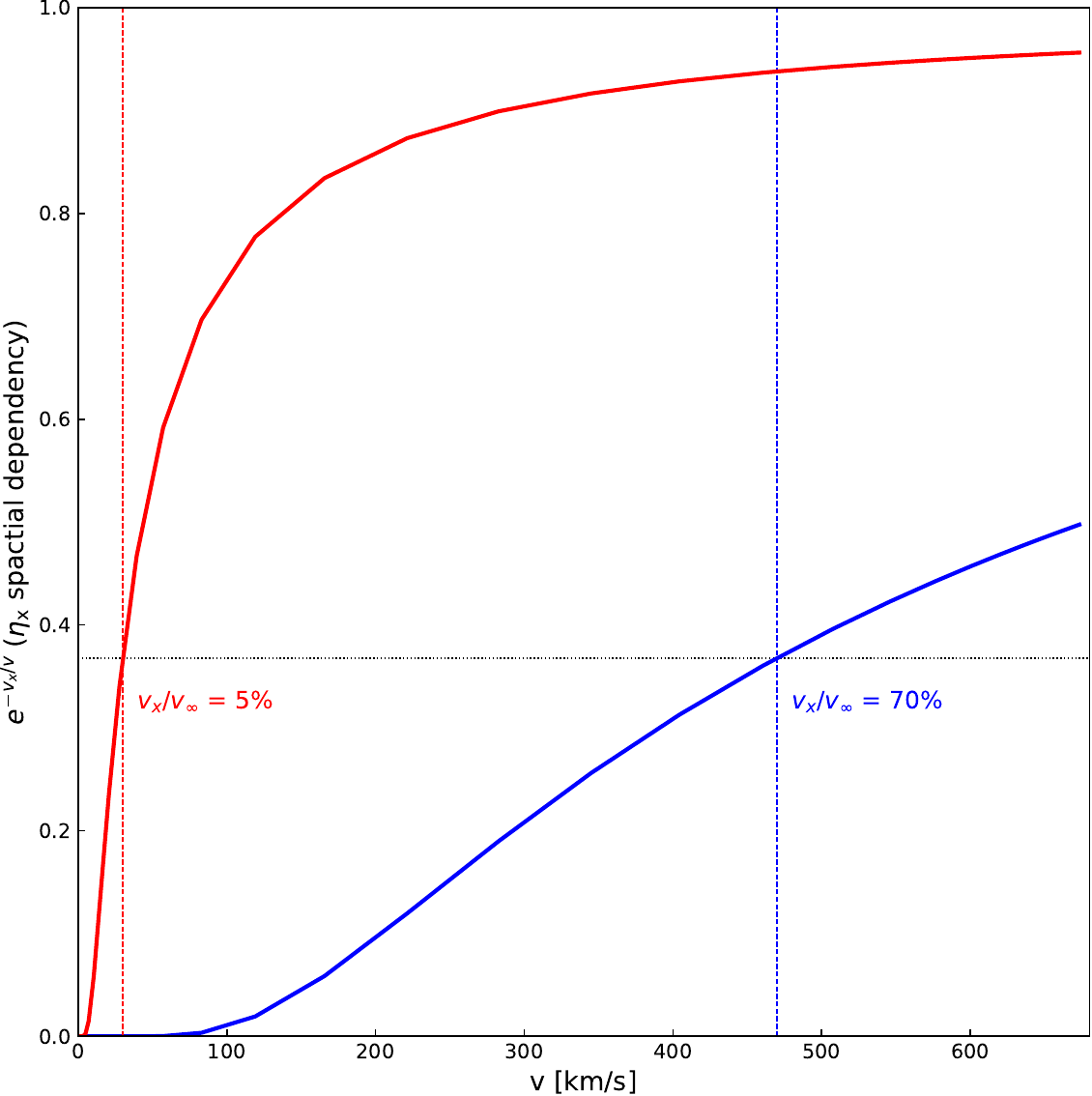}}
  \caption{Velocity dependence of the X-ray emissivity in CMFGEN for different onset velocities: $\varv_\text{x}/\varv_\infty = 5$ \% in red and $\varv_\text{x}/\varv_\infty = 70$ \% in blue. The black dotted horizontal line marks the point where $-\varv_\text{x}/v(r) = 1$, which can be physically interpreted as where the shocks in the wind would roughly become relevant to produce X-rays.}
  \label{fig:exp-vxv}
\end{figure}

In Fig.\,\ref{fig:exp-vxv}, we show an example of the velocity-dependence of $\eta_\text{x}$. A low value of $\varv_\text{x}$ describes a wind where the X-ray emissivity is essentially ubiquitous in the wind. In contrast, a high $\varv_\text{x}$ severely reduces the emissivity in the innermost wind regions (i.e., at low velocities) and describes a wind where X-rays are progressively getting more relevant towards the outermost regions.

\begin{figure}
  \resizebox{\hsize}{!}{\includegraphics{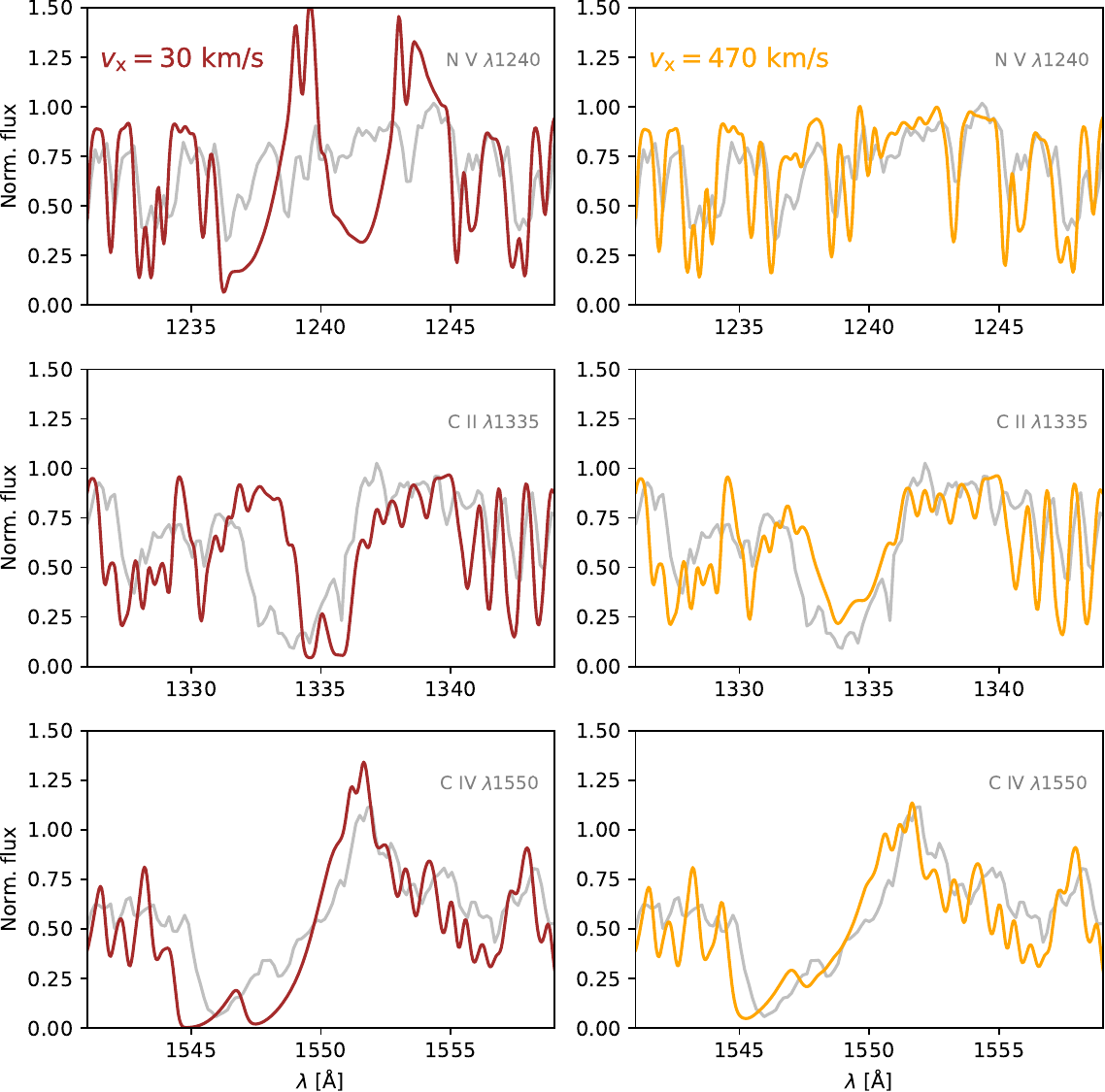}}
  \caption{Comparison between cool BSG models with low and high $\varv_{\mathrm{x}}/\varv_\infty$ respectively (left and right columns of panels respectively).}
  \label{fig:impact-exp-vxv}
\end{figure}

Moreover, Fig.\,\ref{fig:impact-exp-vxv} illustrates that the resulting $L_\text{x}$ is affected by the choice of $\varv_\text{x}$ as the model with high-$\varv_\text{x}$ yields an overall lower factor than the low-$\varv_x$. Thus, a model with a higher $\varv_\text{x}$ parameter also requires a higher filling factor $f_\mathrm{x}$ in order to yield the same X-ray luminosity.

\subsection{CMFGEN vs. PoWR descriptions of X-rays}

As mentioned in Sect.\,\ref{sec:modelling}, PoWR includes X-rays as thermal \textit{Bremsstrahlung} emission similarly to CMFGEN. The resulting expression for $\eta_\text{x}$ is expressed as:
\begin{equation}
    \label{eq:XR_powr}
        \eta_\text{x}(r,\nu) \propto f_\text{x}^2 \, \, \, H(r;r_0) \, \, \, \frac{e^{-h\nu/k_\text{B} T_\text{x}}}{\sqrt{T_\text{x}}} \sum_j^{\mathrm{all \, ions}} z_j^2 g_\text{ff}(\nu,r) n_\text{e}(r) n_j(r),
\end{equation}
where the repeated symbols from Eq.\,\eqref{eq:XR_cmfgen} have the same meaning. A major difference is the spatial distribution of the emissivity, which is modeled by a Heaviside step function\footnote{If $r < r_0$ then $H = 0$ and if $r \geq r_0$ then $H = 1$.} centered in $r_0$ $H(r;r_0)$ in PoWR. This describes a situation where the X-rays start sharply in the wind beyond a certain radius (or velocity), which differs from the gradual build-up in CMFGEN. 

Given the different implementations, transitions affected by X-rays could produce slightly different UV line profile shapes despite having similar $L_\mathrm{x}$, $T_\mathrm{x}$ and ``onset'' descriptions (i.e.,  $\varv_\mathrm{x}$ or corresponding radius). In this study, our reported parameters are all taken from CMFGEN and solely employ PoWR to study potential impacts of optically thick clumping, so the difference in the descriptions is not relevant. Nonetheless, it implies that X-ray luminosities derived only from UV line profiles have a non-negligible dependence on the employed atmosphere code. Detailed observations and multi-dimensional hydrodynamic simulations could prove helpful in gaining quantitative insights on the spatial origin of wind-intrinsic X-rays in order to improve the parametrization in atmosphere models.

\section{SEDs, Optical and UV spectra}
\label{app:master_plots}

\begin{figure*}
\centering
   \includegraphics[width=17cm]{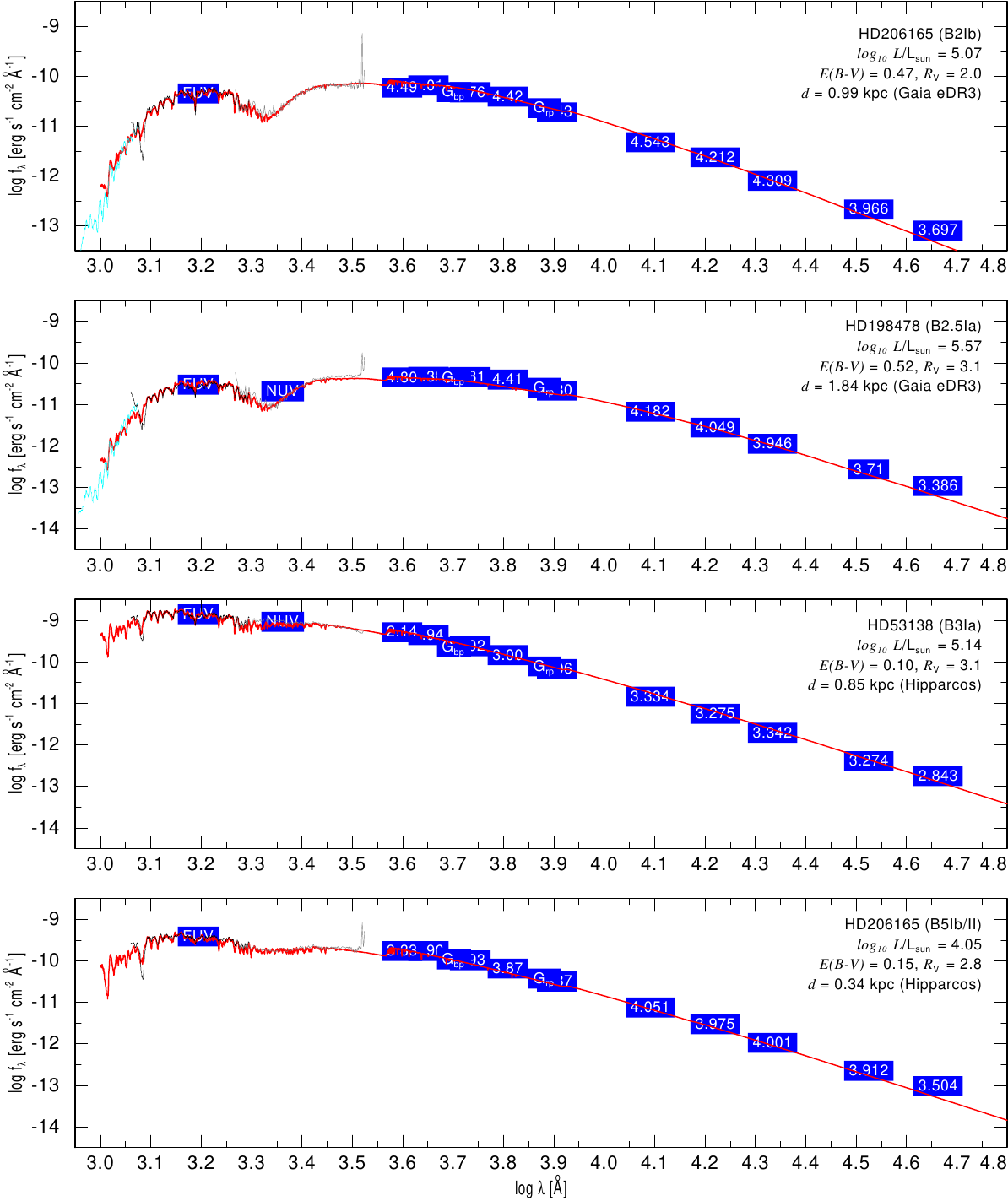}
     \caption{Spectral Energy Distributions of the sample cool BSGs.}
     \label{fig:SED}
\end{figure*}

\begin{figure*}
\centering
   \includegraphics[width=17cm]{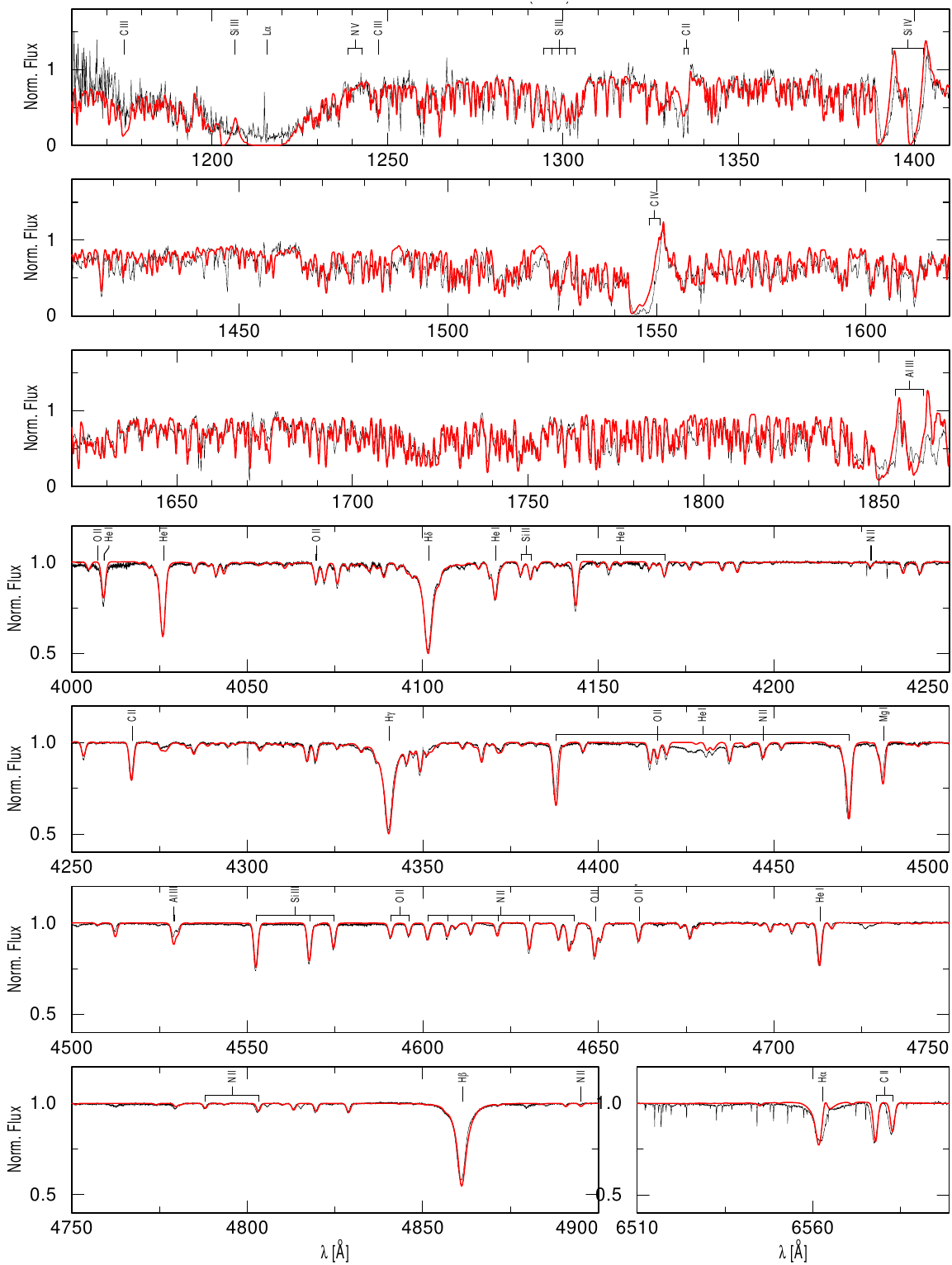}
     \caption{UV and optical spectra of HD206165.}
     \label{fig:HD206165_spec}
\end{figure*}

\begin{figure*}
\centering
   \includegraphics[width=17cm]{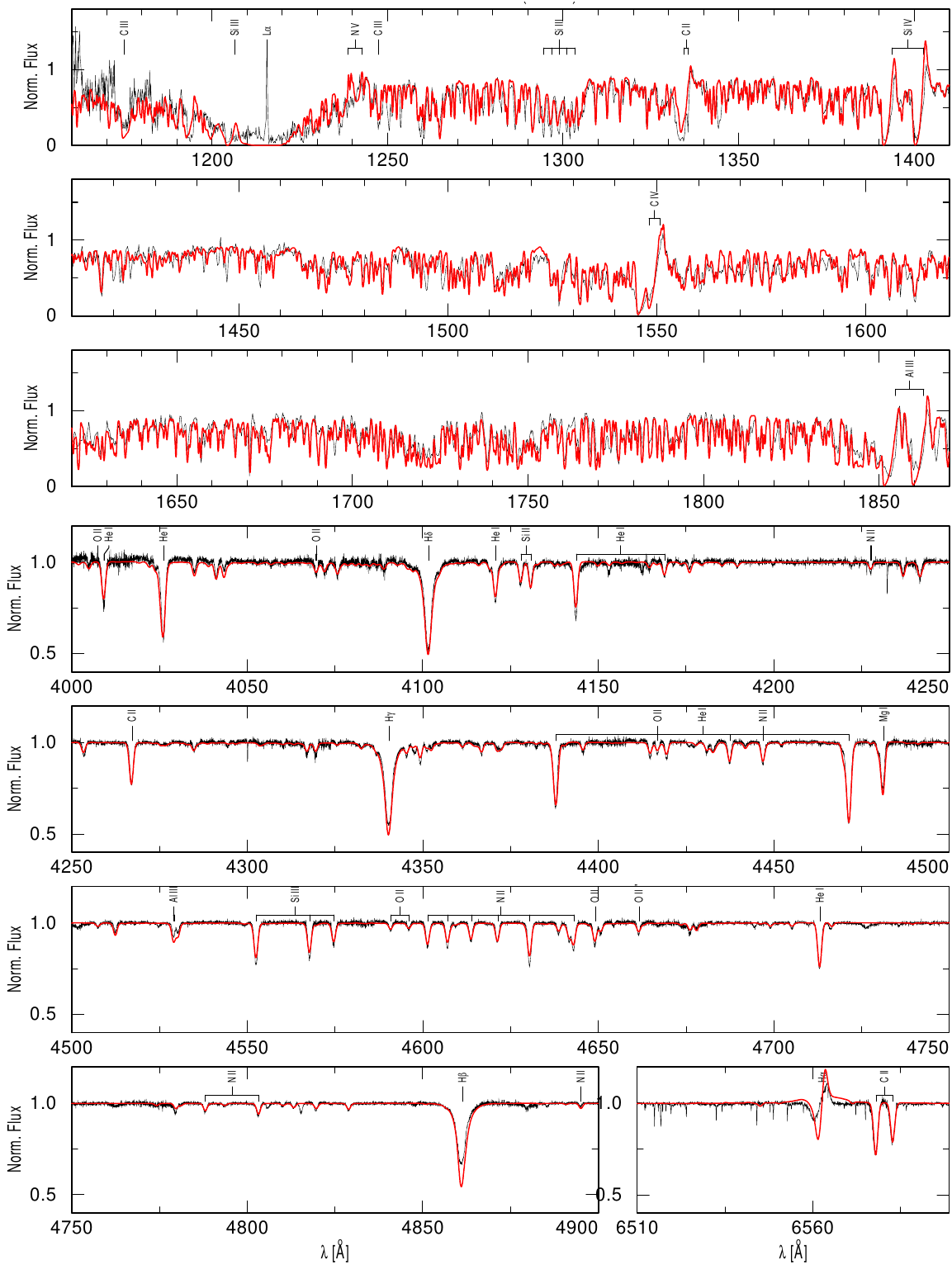}
     \caption{UV and optical spectra of HD198478.}
     \label{fig:HD198478_spec}
\end{figure*}

\begin{figure*}
\centering
   \includegraphics[width=17cm]{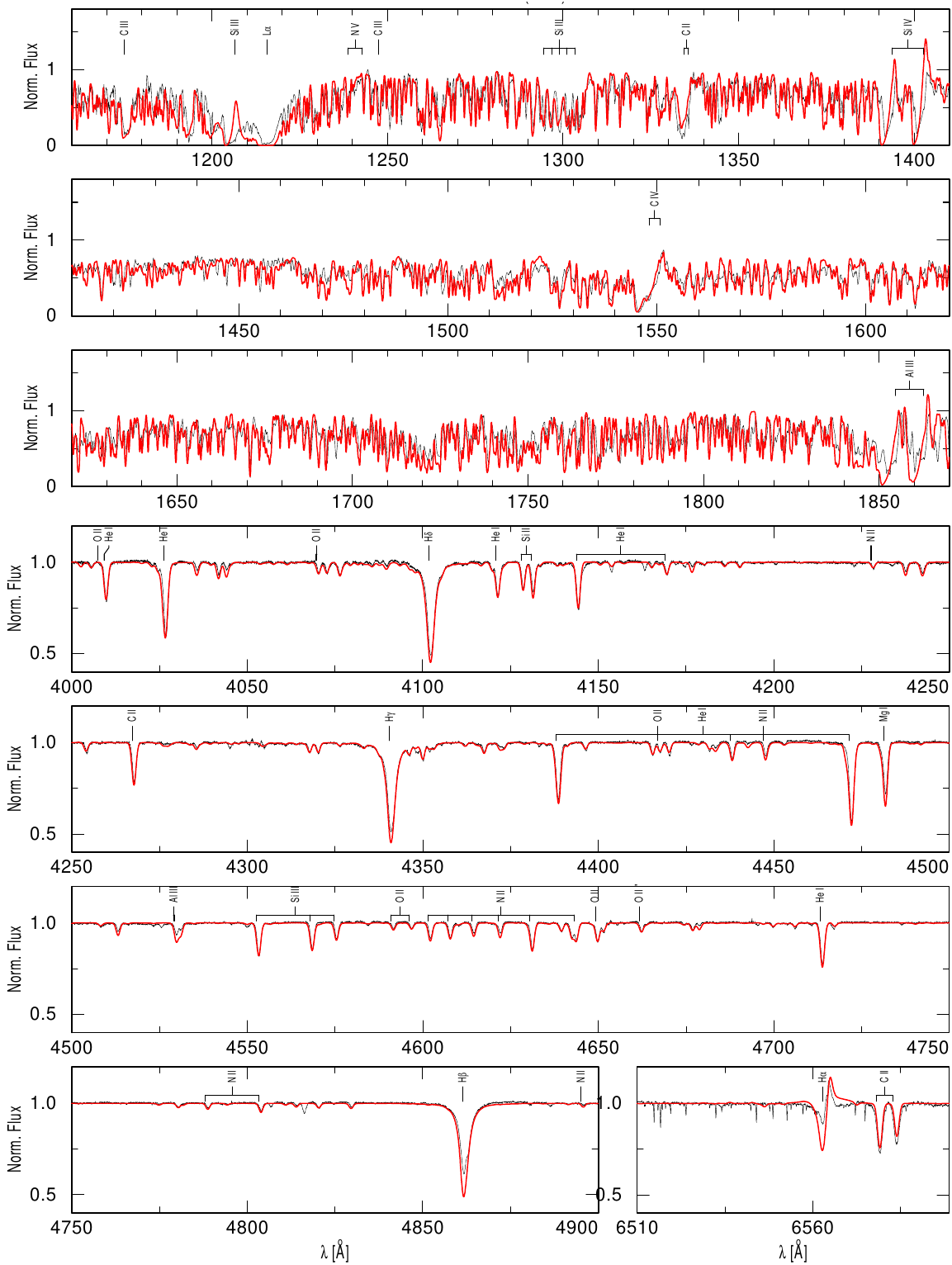}
     \caption{UV and optical spectra of HD53138.}
     \label{fig:HD53138_spec}
\end{figure*}

\begin{figure*}
\centering
   \includegraphics[width=17cm]{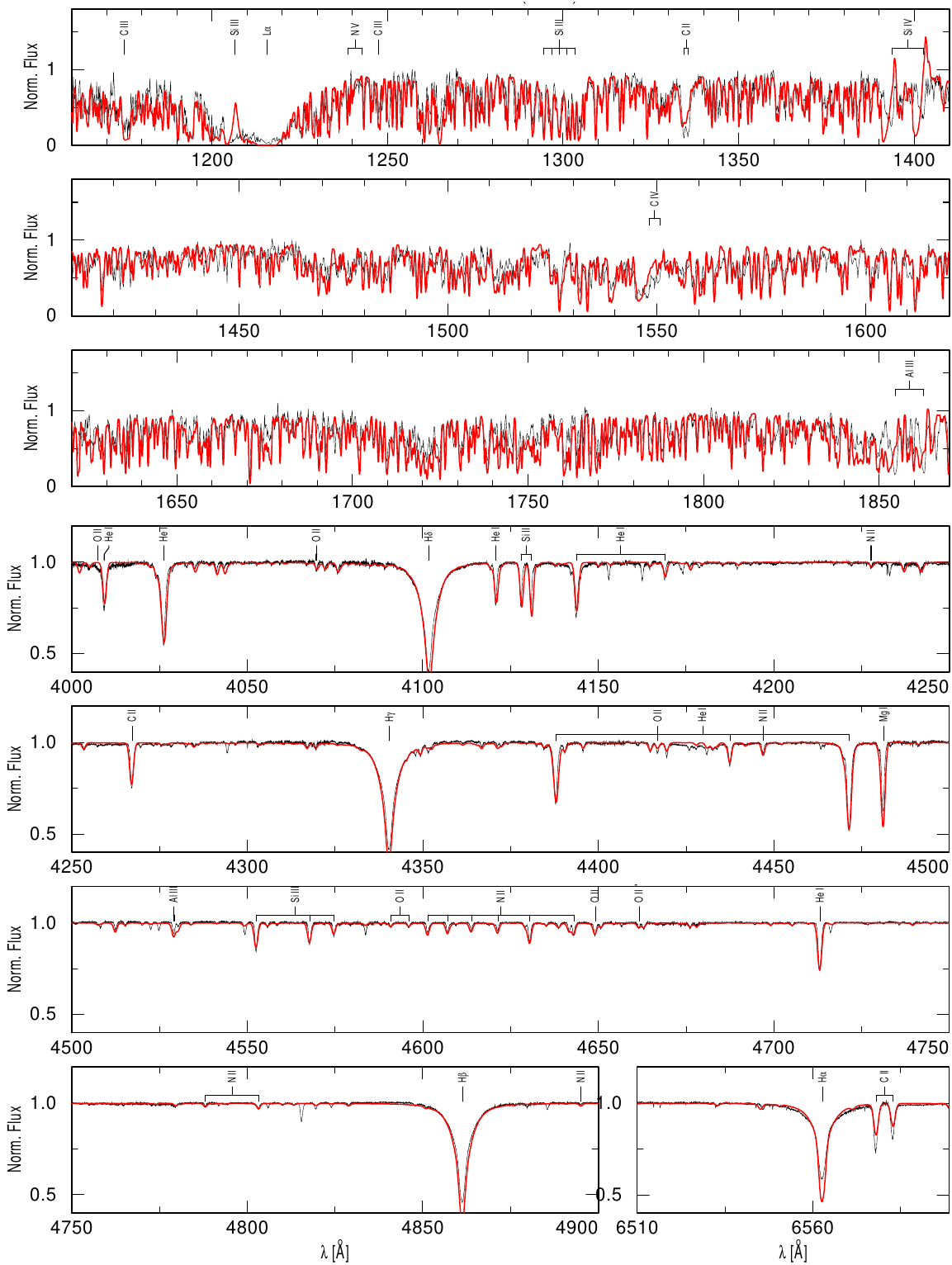}
     \caption{UV and optical spectra of HD164353.}
     \label{fig:HD164353_spec}
\end{figure*}

In this section, we present combined figures showing the spectral energy distribution (SED) and the normalized spectra. In Fig.\,\ref{fig:SED}, the different panels show the different targets studied in this work. The model SEDs are compared to observed flux-calibrated UV spectra (IUE and if available also FUSE) as well as the flux values obtained from the available magnitudes spanning from the UV to the near IR (see Table~\ref{tab:photometry} and Sect.\,\ref{sec:observation} for details). Then, Figs.\,\ref{fig:HD206165_spec} to \ref{fig:HD164353_spec} show the normalized observations in the UV and optical range compared to the best-fitting model spectra.

\begin{table}
\caption{\label{tab:photometry}Magnitudes of the sample BSGs.}
\centering
\begin{tabular}{c|cccc}
\hline
\hline
Stars &  HD206165                    & HD198478                     & HD53138                     &  HD164353                    \\
\hline
FUV   & 7.566                        & 8.020                         & 3.842                       & 5.240                         \\
NUV   &   --                           & 7.599                        & 3.459                       &    --                          \\
U     & 4.490                         & 4.800                          & 2.140                        & 3.330                         \\
B     & 5.006                        & 5.381                        & 2.943                       & 3.959                        \\
V     & 4.760                         & 4.810                         & 3.020                        & 3.930                         \\
R     & 4.420                         & 4.410                         & 3.000                           & 3.870                         \\
Gbp   & 4.824                        & 4.914                        & 3.143                       & 3.955                        \\
G     & 4.655                        & 4.673                        & 3.041                       & 3.932                        \\
Grp   & 4.357                        & 4.273                        & 3.044                       & 3.836                        \\
I     & 4.431                        & 4.301                        & 3.056                       & 3.874                        \\
J     & 4.543                        & 4.182                        & 3.334                       & 4.051                        \\
H     & 4.212                        & 4.049                        & 3.275                       & 3.975                        \\
K     & 4.309                        & 3.946                        & 3.342                       & 4.001                        \\
W1    & 3.966                        & 3.710                         & 3.274                       & 3.912                        \\
W2    & 3.697                        & 3.386                        & 2.843                       & 3.504                        \\

\hline
\end{tabular}
\end{table}

\end{document}